\documentclass[
 reprint,
 amsmath,
 amssymb,
 aps,
 prd,
 preprintnumbers,
 nofootinbib
]{revtex4-1}

\usepackage{slashed}
\usepackage{dsfont}
\usepackage{amsmath}
\usepackage{graphicx}
\usepackage[caption=false]{subfig}
\usepackage{xspace}
\usepackage{xcolor}

\newcommand{\I}{\mathds{1}}
\newcommand{\Z}{\mathds{Z}}

\newcommand{\Tr}{\mathrm{Tr}}
\newcommand{\SU}{\mathrm{SU}}

\newcommand{\Zc}{\mathrm{Z}}

\newcommand{\muhat}{\hat{\mu}}
\newcommand{\nuhat}{\hat{\nu}}
\newcommand{\munu}{{\mu\nu}}
\newcommand{\eq}[1]{(\ref{{1}})}

\begin{document}

\title{Influence of center vortices on the overlap quark propagator in dynamical QCD}

\author{Adam Virgili} \author{Waseem Kamleh} \author{Derek B. Leinweber}
\affiliation{Special Research Centre for the Subatomic Structure of
 	Matter (CSSM), Department of Physics, University of Adelaide,
 	South Australia 5005, Australia.}

\begin{abstract}
There is strong evidence supporting center vortices as underpinning confinement and dynamical chiral symmetry breaking --  the two key features of nonperturbative QCD. In our recent letter~\cite{Kamleh:2023gho} we find that dynamical mass generation vanishes upon vortex removal in full QCD with a near-physical quark mass. In this work we extend those results and consider the influence of center vortex removal on the overlap Landau-gauge quark propagator at multiple valence quark masses on the same dynamical QCD ensemble, keeping fixed the near-physical sea quark mass. After carefully applying a smoothing process we also find that dynamical mass generation is reproduced on the corresponding vortex-only fields. This vortex-only dynamical mass shows qualitative agreement with the untouched Monte Carlo gauge field results. The results reported herein lend further credence to the important relationship between dynamical fermions and center vortices and the mediation of nonperturbative phenomena in QCD.
\end{abstract}

\preprint{ADP-24-15/T1254}

\maketitle

\section{Introduction}

There is a strong and increasing body of evidence supporting centre vortices as the mechanism underpinning confinement and dynamical chiral symmetry breaking, the two key features of low-energy, nonperturbative QCD~\cite{tHooft:1977nqb,tHooft:1979rtg,DelDebbio:1996lih,Faber:1997rp,DelDebbio:1998luz,Bertle:1999tw,deForcrand:1999our,Faber:1999gu,Engelhardt:1999fd,Engelhardt:1999xw,Engelhardt:2000wc,Bertle:2000qv,Langfeld:2001cz,Engelhardt:2002qs,Langfeld:2003ev,Greensite:2003bk,Bruckmann:2003yd,Engelhardt:2003wm,Boyko:2006ic,Ilgenfritz:2007ua,Bornyakov:2007fz,Bowman:2008qd,Hollwieser:2008tq,OCais:2008kqh,Engelhardt:2010ft,Bowman:2010zr,OMalley:2011aa,Hollwieser:2013xja,Hollwieser:2014soz,Trewartha:2015ida,Trewartha:2015nna,Greensite:2016pfc,Trewartha:2017ive,Biddle:2018dtc,Spengler:2018dxt,Biddle:2022zgw,Biddle:2022acd}.
Whilst the results of these studies demonstrate a clear role for centre vortices in the mediation of these phenomena, it is also clear there is more to be understood.

This is apparent in recent studies of the static quark potential and the gluon propagator in dynamical QCD.
Whilst vortex removal in pure-$\SU(3)$ gauge results in a loss of string tension, and thus confinement, only about two-thirds of the original string tension is recovered on the corresponding vortex-only background~\cite{OCais:2008kqh,Trewartha:2015ida,Langfeld:2003ev}.
The loss of string tension upon vortex removal is a clear sign of the important role of centre vortices in confinement, and is further supported by its partial recovery on a vortex-only background.
However, the `missing' third of the string tension remains unaccounted for, suggesting there is more to be understood.
This quantitative discrepancy is broadly illustrative of the findings of pure-gauge centre-vortex studies, some of which are discussed in further detail below.

In light of these results, a recent study~\cite{Biddle:2022zgw} explored the impact of dynamical fermions on the relationship between centre vortices and the string tension.
Here, the centre vortex structure of configurations generated with 2+1 dynamical flavours of fermion~\cite{PACS-CS:2008bkb} were examined.
Once again, in agreement with pure-gauge results, the string tension vanishes upon vortex removal.
However, it was found that the original string tension is fully recovered on a vortex-only background.
The quantitative discrepancy present in the pure-gauge sector is resolved in the presence of dynamical fermions.

A similar pattern is observed in studies of the gluon propagator.
Although the infrared enhancement of the gluon propagator is suppressed upon vortex removal in pure gauge, a residual strength still persists~\cite{Bowman:2010zr,Biddle:2018dtc,Biddle:2022acd}.
This is consistent with positivity violation observed in the correlator at large distances, and demonstrates an imperfect separation between perturbative and nonperturbative physics upon vortex modification.
In the presence of dynamical fermions, however, the vortex-removed correlator is consistent with positivity, and the residual infrared strength in the propagator is significantly diminished~\cite{Biddle:2022acd}.
As such, vortex modification in the presence of dynamical fermions provides an effective separation between the perturbative and nonperturbative aspects of QCD. 

These results demonstrate an intimate link between dynamical fermions and centre vortices. Although the nature of this relationship is at this stage unclear, the resolution of the quantitative discrepancies present in the pure-gauge sector provide further strong evidence supporting centre vortices as the origin of confinement in QCD. 
The existence of such a link provides strong motivation to consider the effect of dynamical fermions on the role of centre vortices in dynamical chiral symmetry breaking.

It is well-established that centre vortices are responsible for dynamical chiral symmetry breaking in $\mathrm{SU}(2)$ gauge theory~\cite{deForcrand:1999our,Engelhardt:2002qs,Bornyakov:2007fz,Hollwieser:2008tq,Bowman:2008qd,Hollwieser:2013xja,Hollwieser:2014soz}.
Meanwhile, hadron spectra computed with Wilson fermions on pure-$\SU(3)$ gauge vortex-removed ensembles display an absence of dynamical chiral symmetry breaking~\cite{OMalley:2011aa}.
A similar study which instead employed overlap fermions was not only able to demonstrate the restoration of chiral symmetry upon vortex-removal, but reproduced the salient features of the spectrum on a vortex-only background~\cite{Trewartha:2017ive}.
There are, however, discrepancies in the masses obtained on the vortex-only background which are lower than those obtained from the original, untouched ensemble, although this may simply be an artefact of the cooling applied to the vortex-only gauge fields.

Less readily explainable discrepancies are manifest in the pure-$\SU(3)$ gauge Landau-gauge overlap quark propagator~\cite{Trewartha:2015nna}.
Although dynamical mass generation in the quark propagator mass function is fully reproduced on a vortex-only ensemble, a clear signal of dynamical chiral symmetry breaking, a quantitative discrepancy arises in the persistence of dynamical mass generation upon vortex-removal.
This once again suggests the presence of pure-gauge sector artifacts in the identified centre vortex fields.

By contrast our recent letter~\cite{Kamleh:2023gho} found that dynamical mass generation vanishes upon vortex removal in full QCD with a near-physical quark mass. We extend those results herein by investigating the behaviour of the Landau-gauge overlap quark propagator upon vortex modification in dynamical QCD at multiple valence quark masses (keeping the sea quark sector fixed). In addition to the vortex-removed quark propagator, we also study the vortex-only propagator in order to perform a comparison with the untouched fields. As has been noted elsewhere~\cite{Virgili:2022ybm}, smoothing of the vortex-only fields is required to achieve locality for the overlap fermion action.

The paper is structured as follows.
Section~\ref{sec:centre_vortices} introduces center vortices, and describes the methods of their identification and removal.
Section~\ref{sec:quark_prop} outlines the Landau-gauge overlap quark propagator, with discussions on overlap fermions (Subsection~\ref{subsec:overlap}), the extraction of the mass and renormalisation functions (Subsection~\ref{subsec:quark_prop}), and Landau gauge fixing (Subsection~\ref{subsec:landau_gauge}).
Section~\ref{sec:vortex-removed} outlines the parameters of the vortex-removed calculation of the Landau-gauge overlap quark propagator and presents results. 
Section~\ref{sec:vortex-only_quark_propagator} does likewise for initial vortex-only calculations of the Landau-gauge quark propagator, each employing a different method of smoothing the rough vortex-only gauge field configurations.
Section~\ref{sec:profile_fitting} continues this comparison of smoothing approaches by considering the instanton-like profiles of the respective gauge fields obtained via each method.
In Section~\ref{sec:vortex-only_comparison}, a single smoothing approach is chosen and the Landau-gauge overlap quark propagator is computed with improved statistics.
Conclusions are discussed in Section~\ref{sec:conclusions}.

\section{Center vortices}
\label{sec:centre_vortices}

On the lattice, center vortices are revealed by projecting each gauge link to an element of the center $\Zc(N)$ of $\SU(N)$ where
\begin{equation}
    \begin{aligned}
        \mathrm{Z}(N) &\equiv \{ g \in \SU(N)\, \vert\, gh = hg\, \forall\, h \in \SU(N) \} \\ 
        &= \{ e^{i 2 \pi n / N} \I\, \vert\, n \in \Z_N \}
    \end{aligned}
\end{equation}
is the set of elements in $\SU(N)$ which commute with every other element of the group.

To obtain the center-projected links, the gauge field is first fixed to maximal center gauge (MCG) by choosing the gauge transform $U_\mu(x) \to U^G_\mu(x)$ which maximizes the functional~\cite{Langfeld:2003ev}
\begin{equation}
  \sum_{x,\mu}\left|{\Tr \, U^G_\mu(x)} \right|^2 \, ,
  \label{eq:MCG}
\end{equation}
as outlined in Refs.~\cite{Montero:1999by,Faber:1999sq,Biddle:2022zgw}. 
Each link is then projected to the nearest element of $\Zc(3)$, such that
\begin{equation}
    U^G_\mu(x) \to \mathcal{P}_{\Zc(3)} \left\{ U^G_\mu(x) \right\} \equiv Z_\mu(x) = e^{i\frac{2\pi}{3}n_\mu(x)}\I
    \label{eq:centre_project}
\end{equation}
where
\begin{equation}
    n_\mu(x) =
    \begin{cases}
        &\phantom{+}0, \text{ if } \arg\Tr\,U^G_\mu(x) \in \left(-\frac{\pi}{3}\,,+\frac{\pi}{3}\right) \,, \\
        &+1, \text{ if } \arg\Tr\,U^G_\mu(x) \in \left(+\frac{\pi}{3}\,,+\pi\right) \,, \\
        &-1, \text{ if } \arg\Tr\,U^G_\mu(x) \in \left(-\pi\,,-\frac{\pi}{3}\right) \,.
    \end{cases}
    \label{eq:centre_element}
\end{equation}
The projected links $Z_\mu(x)$ define a center-vortex configuration in MCG. 
The elementary plaquette $P_{\mu\nu}(x)$ is given by the product of links $U$ around a unit square, 
\begin{equation}
  P_{\mu\nu}(x) = U_\mu(x)\,U_\nu(x+\muhat)\,U_\mu^\dagger(x+\nuhat)\,U_\nu^\dagger(x).
\end{equation}
Center vortices are identified by the vortex flux through each vortex-projected plaquette, where
\begin{equation}
    \begin{aligned}
        P_{\mu\nu}(x) &= Z_\mu(x)\,Z_\nu(x+\muhat)\,Z_\mu^\dagger(x+\nuhat)\,Z_\nu^\dagger(x)\\ 
        &= e^{i\frac{2\pi}{3}p_\munu(x)}\I \, ,
    \end{aligned}
\end{equation}
corresponds to a vortex flux value $p_\munu(x) \in \left\{-1,0,1\right\}$. 
A plaquette with vortex flux $p_\munu(x)=\pm1$ is identified as pierced by a vortex with center charge $\pm1$.
A vortex-removed link $U^{\mathrm{VR}}_\mu(x)$ is simply the product of the MCG-fixed link and the inverse of its center-projected link, given by
\begin{equation}
    U^{\mathrm{VR}}_\mu(x) = Z_\mu^\dagger(x) \, U_\mu^G(x)\,,
\end{equation}
and defines a vortex-removed configuration in MCG.

Throughout this work we refer to the original $\SU(3)$ gauge fields as {\it{untouched}} and the gauge fields generated by vortex removal as {\it{vortex-removed}}.

\section{Landau-gauge overlap quark propagator}
\label{sec:quark_prop}

\subsection{Overlap fermions}
\label{subsec:overlap}

Within the overlap formalism~\cite{Narayanan:1993zzh,Narayanan:1993sk,Narayanan:1993ss,Narayanan:1994gw,Neuberger:1997fp,Kikukawa:1997qh}, the massless overlap Dirac operator is given by
\begin{equation}
    D_{o} = \frac{1}{2a} \left( 1 + \gamma^5 \epsilon \left(H\right) \right)\,,
\end{equation}
where $\epsilon (H)$ is the matrix sign function applied to the overlap kernel $H$.
Typically, the kernel is chosen to be the Hermitian Wilson Dirac operator, but other choices are valid and in particular the use of a kernel which incorporates smearing can have numerical advantages~\cite{Kamleh:2001ff,Bietenholz:2002ks,Kovacs:2002nz,DeGrand:2004nq,Durr:2005mq,Durr:2005ik,Bietenholz:2006fj}.
In this work we employ two different kernels.

For the calculation of the quark propagator on a vortex-removed background, we employ the fat-link irrelevant clover (FLIC) fermion action~\cite{Zanotti:2001yb,Kamleh:2001ff,Kamleh:2004xk,Kamleh:2004aw} $H=\gamma^5D_\text{flic}$ where
\begin{equation}
	D_\text{flic} = \slashed{\nabla}_\text{mfi} + \frac{a}{2} \left( \Delta^\text{fl}_\text{mfi} - \frac{1}{2}\sigma\cdot F^\text{fl}_\text{mfi} \right) + m_\text{w} \,,
\end{equation}
consistent with the untouched calculation of Ref.~\cite{Virgili:2022wfx}.
The Wilson hopping parameter $\kappa$ is related to $m_\text{w}$ by
\begin{equation}
	\kappa \equiv \frac{1}{8-2am_\text{w}}\,.
\end{equation} 
The subscript ``mfi'' in the FLIC kernel denotes the use of gauge links which have been mean-field improved~\cite{LePage:1992xa} by taking
\begin{equation}
    U_\mu(x) \to \frac{U_\mu(x)}{u_0}
\end{equation}
where
\begin{equation}
    u_0 = \langle \frac{1}{3} \Re \Tr \left[ P_\munu(x) \right] \rangle^{\frac{1}{4}}
\end{equation}
is the mean link.
Furthermore, the Wilson and clover terms of the FLIC kernel are constructed from {\it{fat links}}, denoted by the superscript ``fl'', which have undergone four sweeps of stout-link smearing~\cite{Morningstar:2003gk} at $\rho=0.1$.
For these terms, mean-field improvement is applied to the fat links.
The utility of the FLIC kernel is the significant improvement of the condition number following the projection of the low-lying eigenmodes.

On the other hand, as the vortex-only background has already undergone smoothing (necessary to satisfy the smoothness condition of the overlap operator) any additional smoothing provided by the kernel is redundant, offering little-to-no benefit.
Furthermore, the mean link of these smoothed gauge fields $u_0 \approx 1.00$, and as such mean-field improvement is also made redundant.
Without these, the FLIC action reduces to the clover fermion action~\cite{Sheikholeslami:1985ij}
\begin{equation}
	D_\text{clover} = \slashed{\nabla} + \frac{a}{2} \left( \Delta - \frac{1}{2}\sigma\cdot F\right) + m_\text{w}\,,
\end{equation}
on which it is based.
We therefore choose $H=\gamma^5D_\text{clover}$ for the vortex-only calculations of the quark propagtor.

\subsection{Quark propagator}
\label{subsec:quark_prop}

The massive overlap Dirac operator~\cite{Neuberger:1997bg} is defined as
\begin{equation}
    D_{o}(\mu) = (1-\mu)D_o + \mu\,,
\end{equation}
where $0 \le \mu \le 1$ is the overlap fermion mass parameter related to the bare quark mass by
\begin{equation}
	m_q = 2m_\text{w}\,\mu\,.
\end{equation}
The calculation of the external overlap fermion propagator requires the subtraction of a contact term. After solving the linear system for a given fermion source $\psi,$
\begin{equation}
D_o(\mu)\chi = \psi \, ,
\end{equation}
each solution vector is modified as
\begin{equation}
\chi_{c} \equiv \frac{1}{2m_{\rm w}(1-\mu)}(\chi - \psi) \, ,
\end{equation}
in order to construct the external overlap quark propagator~\cite{Narayanan:1994gw,Edwards:1998wx},
\begin{equation}
    S(p) \equiv \frac{1}{2m_{\rm w}(1-\mu)}(D^{-1}_o(\mu) - 1) \, .
\end{equation}
Defining the bare mass $m^0$ via
\begin{equation}
m^0 = 2m_{\rm w}\mu \, ,
\end{equation}
through the above subtraction of the contact term, it is possible to show that
\begin{equation}
    S^{-1}(p) = S^{-1}(p) \vert_{m^0 = 0} + m^0 \, . 
\end{equation}
and that exact chiral symmetry is obeyed
\begin{equation}
    \{\gamma^5 ,S(p)\vert_{m^0 = 0} \} = 0 \, ,
\end{equation}
just as in the continuum~\cite{Neuberger:1997fp}. 

The general form of the (color-traced) lattice overlap quark propagator in momentum space can be written as
\begin{equation}
    S(p) = \frac{Z(p)}{i \slashed{q} + M(p)}
    \label{eq:S_ctr}
\end{equation}
where $Z(p)$ is the renormalisation function, $M(p)$ is the mass function, and $q_\mu$ is the kinematical lattice momentum defined by
\begin{equation}
	S_\text{tree}^{-1}(p)=i\slashed{q} + m_\text{w} \,,
\end{equation}
where the subscript ``tree'' denotes use of link variables $U_\mu(x) = \I \ \ \forall \ \ x,\,\mu$.
This is the only tree-level correction required for the overlap quark propagator.
The simple form of Eq.~(\ref{eq:S_ctr}) is afforded by the absence of additive renormalisation in the overlap formalism.
Isolation of $M(p)$ and $Z(p)$ is straight forward.
Rewriting Eq.~(\ref{eq:S_ctr}) as
\begin{equation}
\begin{aligned}
	S_\mathrm{ctr}(p) 
	&= \frac{-i\slashed{q}Z(p) + M(p)Z(p)}{q^2+M^2(p)}
	\equiv -i\slashed{\mathcal{C}}(p) + \mathcal{B}(p)\,,
\end{aligned}
\end{equation}
where we have defined
\begin{align}
    \mathcal{B}(p) &\equiv \frac{1}{n_sn_c}\Tr\left[S(p)\right] = \frac{M(p)Z(p)}{q^2+M^2(p)}\,, \\
    \mathcal{C}_\mu(p) &\equiv \frac{i}{n_sn_c}\Tr\left[\gamma_\mu S(p)\right] = \frac{q_\mu Z(p)}{q^2+M^2(p)}\,,
\end{align}
and $n_s$ and $n_c$ are the extents of spin and color indices.
Defining
\begin{equation}
    \mathcal{A}(p) = \frac{q \cdot \mathcal{C}}{q^2} = \frac{Z(p)}{q^2+M^2(p)} \,,
\end{equation}
the mass and renormalisation functions are isolated by
\begin{align}
    M(p) &= \frac{\mathcal{B}(p)}{\mathcal{A}(p)}\,, \\
    Z(p) &= \frac{\mathcal{C}^2(p) + \mathcal{B}^2(p)}{\mathcal{A}(p)}\,.
\end{align}

\subsection{Landau gauge fixing}
\label{subsec:landau_gauge}

The quark propagator is gauge dependent, and hence requires a choice of gauge fixing condition.
In this work we use the Landau gauge condition, which in the continuum is defined by
\begin{equation}
    \partial_\mu A_\mu(x) = 0\,.
\end{equation}
On the lattice, this condition is satisfied by finding the gauge transformation which maximizes the $\mathcal{O}(a^2)$-improved functional~\cite{Bonnet:1999mj}
\begin{equation}
    \mathcal{F}_\text{Imp} = \frac{4}{3}\mathcal{F}_1 - \frac{1}{12u_0}\mathcal{F}_2\,,
    \label{eq:functional}
\end{equation}
where
\begin{align}
    \mathcal{F}_1 &= \sum_{x,\mu}\frac{1}{2}\Tr\left[U_\mu(x) + U^\dagger_\mu(x)\right]\,, \\
    \mathcal{F}_2 &= \sum_{x,\mu}\frac{1}{2}\Tr\left[U_\mu(x)U_\mu(x+\muhat) + U^\dagger_\mu(x+\muhat)U^\dagger_\mu(x)\right]\,.
\end{align}
We use the Fourier accelerated conjugate gradient method~\cite{Hudspith:2014oja} to optimize Eq.~(\ref{eq:functional}).

\section{Vortex-removed quark propagator}
\label{sec:vortex-removed}

\subsection{Simulation parameters}

We compute the Landau-gauge overlap quark propagator on a vortex-removed $32^3 \times 64$ PACS-CS 2+1 flavor ensemble~\cite{PACS-CS:2008bkb} at the lightest available pion mass $m_\pi = 156$ MeV with lattice spacing $a=0.0933$ fm corresponding to S\"ommer parameter $r_0 = 0.49$ fm.
The large volume of the lattice provides significant averaging over each configuration such that statistically accurate results are obtained on 30 gauge field configurations.
The FLIC overlap fermion action was employed at six quark masses $m_q = 6,\,9,\,19,\,28,\,56,\,84$ MeV where the lightest mass was tuned to match the pion mass of the ensemble.
The Wilson mass parameter was set to $am_\text{w}=-1.1,$ corresponding to a hopping parameter of $\kappa=0.17241$ in the kernel.
The matrix sign function was calculated using the Zolotarev rational polynomial approximation~\cite{Chiu:2002eh}.
The evaluation of the inner conjugate gradient was accelerated by projecting out the 80 lowest-lying eigenmodes and calculating the sign function explicitly.
Finally, a cylinder~\cite{Leinweber:1998im} cut is applied to the propagator data. 
$Z(p)$ is renormalized in the MOM scheme~\cite{Leinweber:1998uu} to be 1 at largest momentum considered $p=6.8$ GeV.
With the exception of vortex-modification, the gauge field ensemble and quark propagator calculations parameters are identical to the untouched propagator of Ref.~\cite{Virgili:2022wfx}.

\subsection{Results}

\begin{figure*}[!ht]
    \centering
    \includegraphics[width=0.9\linewidth]{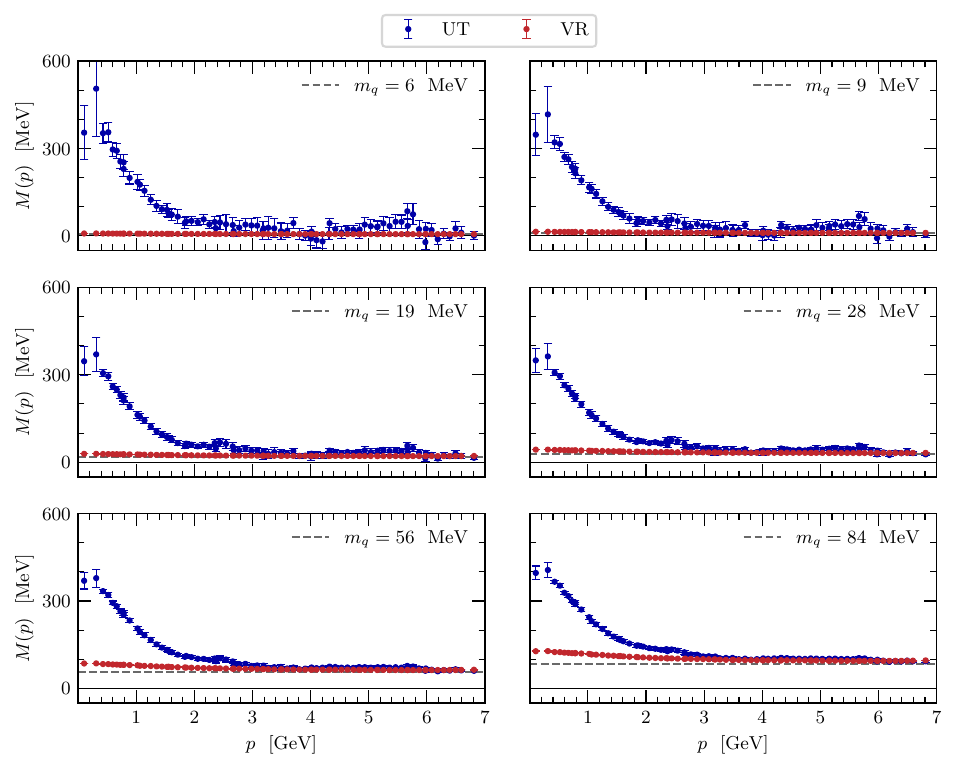}
    \caption{Untouched (blue) and vortex-removed (red) mass functions $M(p)$ for all quark masses considered.}
    \label{fig:M_VR}
\end{figure*}
\begin{figure*}[!ht]
    \centering
    \includegraphics[width=0.9\linewidth]{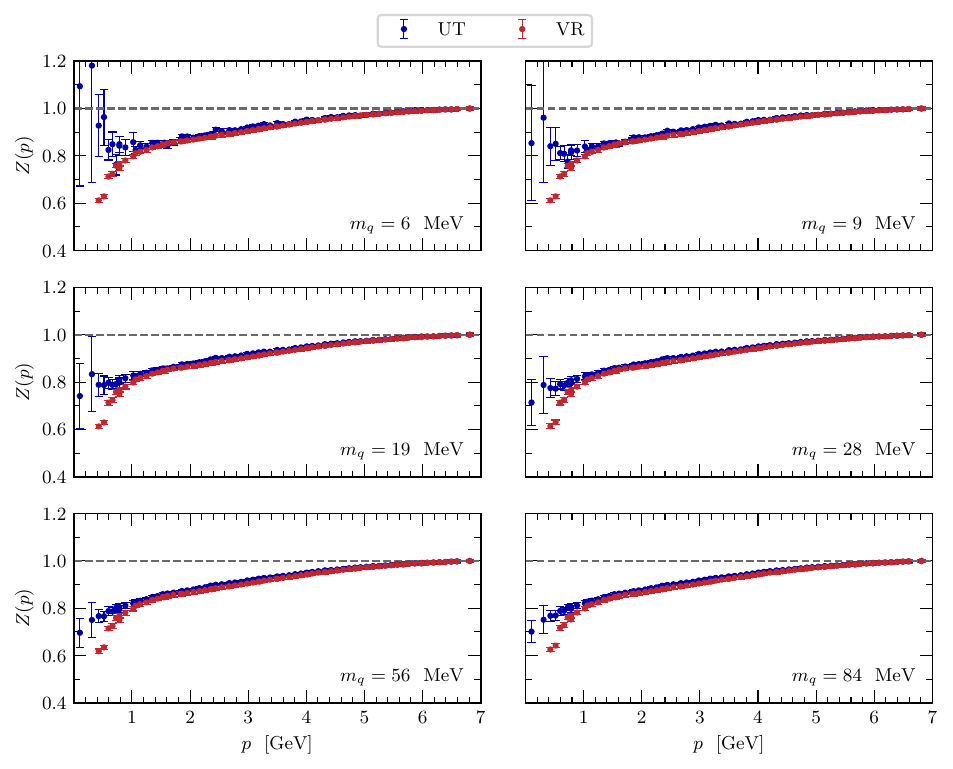}
    \caption{Untouched (blue) and vortex-removed (red) renormalisation functions $Z(p)$ for all masses considered.}
    \label{fig:Z_VR}
\end{figure*}

\begin{figure}[!h]
    \centering
    \subfloat[\label{fig:compareM_pure}]{\includegraphics[width=0.9\linewidth]{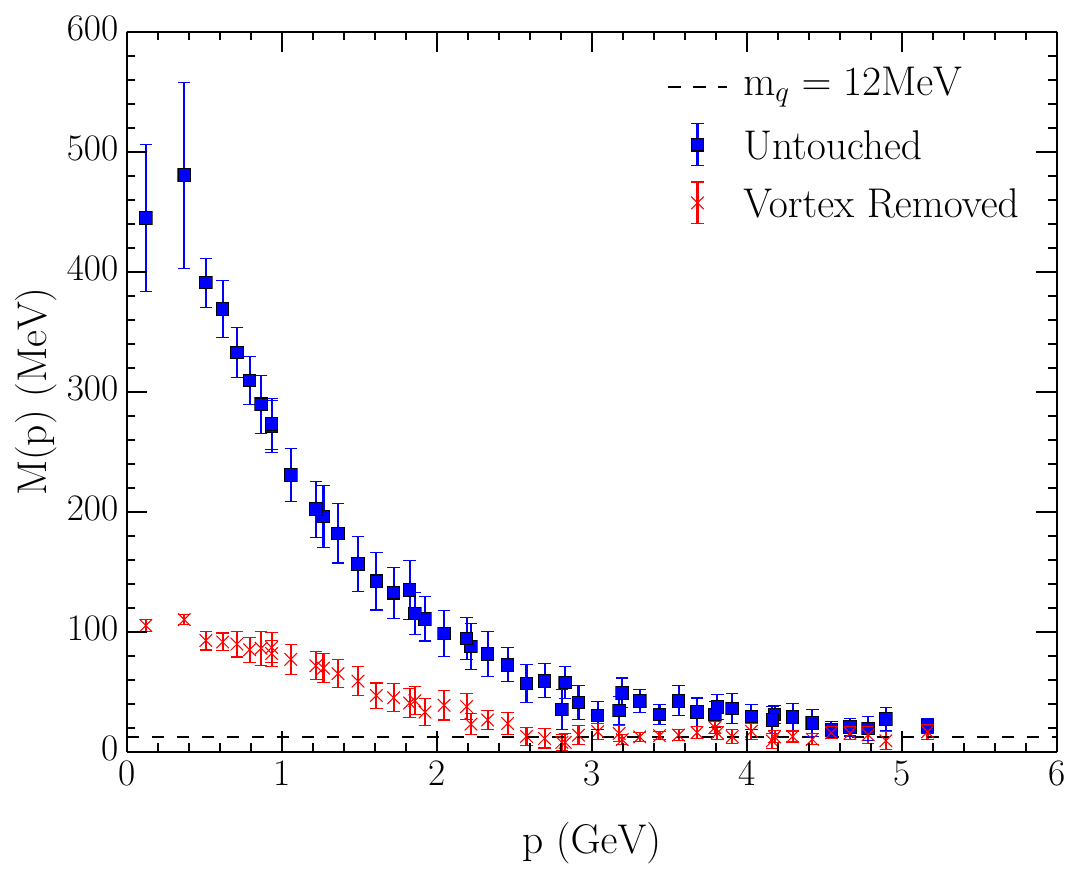}} \\
    \subfloat[]{\includegraphics[width=0.9\linewidth]{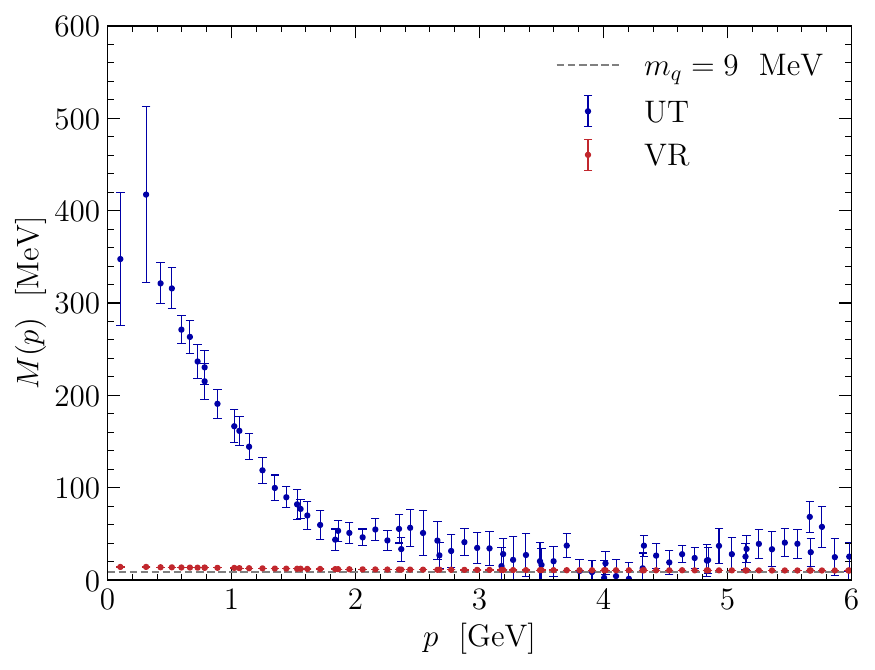}}
    \caption{Comparison of mass functions from Ref.~\cite{Trewartha:2015nna} in pure-gauge for $m_q=12$~MeV (a), and this study for $m_q=9$ MeV (b). While 12 MeV is the lightest available mass from Ref.~\cite{Trewartha:2015nna}, 9 MeV is the closest mass in the current study. Note that the axes of (b) have been adjusted to match (a).}
    \label{fig:compareM}
\end{figure}
\begin{figure}[!h]
    \centering
    \subfloat[\label{fig:compareZ_pure}]{\includegraphics[width=0.9\linewidth]{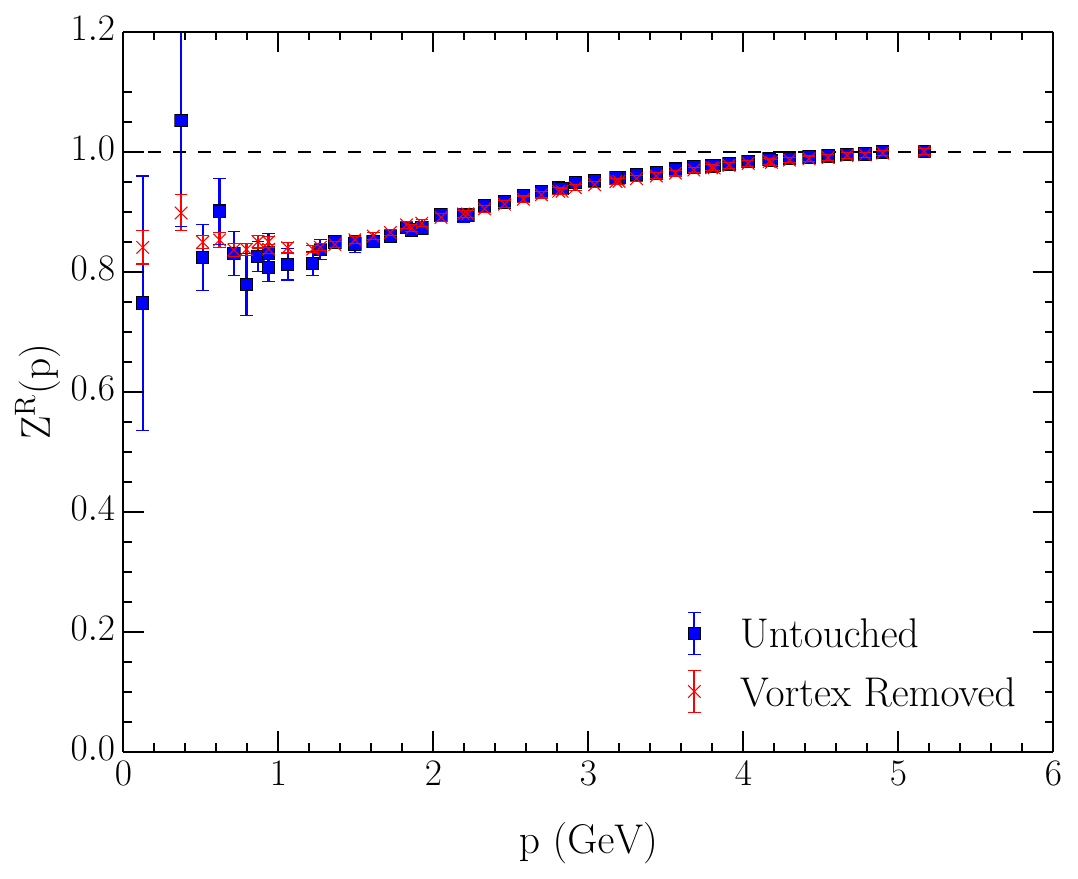}} \\
    \subfloat[]{\includegraphics[width=0.9\linewidth]{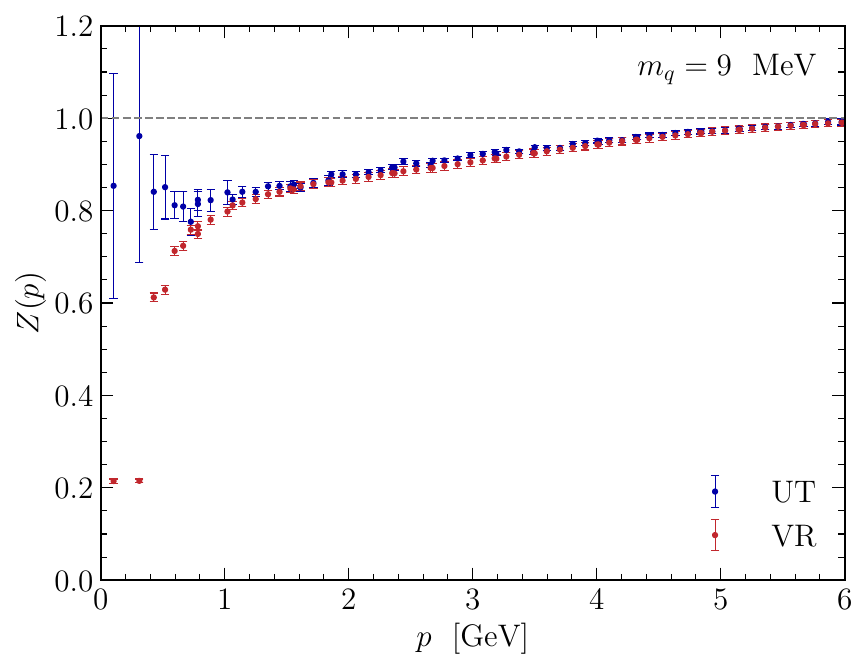}} 
    \caption{Comparison of renormalisation functions from Ref.~\cite{Trewartha:2015nna} in pure-gauge for $m_q=12$~MeV (a), and this study for $m_q=9$ MeV (b). While 12 MeV is the lightest available mass from Ref.~\cite{Trewartha:2015nna}, 9 MeV is the closest mass in the current study. Note that the axes of (b) have been adjusted to match (a).}
    \label{fig:compareZ}
\end{figure}

The vortex-removed mass function is compared with the untouched mass function of Ref.~\cite{Virgili:2022wfx} in Figure~\ref{fig:M_VR}, for each mass considered.
As we reported in Ref.~\cite{Kamleh:2023gho}, upon vortex removal dynamical mass generation vanishes at the lightest quark mass.
At a valence quark mass of $m_q = 19$~MeV we first see the appearance of a slight remnant of dynamical mass generation that then gradually increases with increasing quark mass, consistent with our understanding that the bare quark mass parameter governs the level of explicit chiral symmetry breaking.
This is reflected by the small remnant of dynamical mass generation becoming more prominent with increasing quark mass.
This also offers a framework to interpret the renormalisation function results presented in Figure~\ref{fig:Z_VR}.

We compare the current dynamical results with those from Ref.~\cite{Trewartha:2015nna} which were obtained in the pure-gauge sector in Figures~\ref{fig:compareM}~and~\ref{fig:compareZ}. For this purpose, we select the lightest available quark mass from Ref.~\cite{Trewartha:2015nna}, $m_q=12$ MeV, and compare it to the nearest (and second-lightest) quark mass considered in this work, $m_q=9$ MeV. Whilst dynamical mass generation is suppressed upon vortex removal in the pure-gauge sector, a significant remnant with a peak above 100 MeV persists as seen in Figure~\ref{fig:compareM_pure}. In the dynamical case, however, the removal of centre vortices suppresses dynamical mass generation to the fullest extent possible in the presence of some small explicit chiral symmetry breaking associated with the finite bare quark mass. This resolves an important discrepancy that we observe in the pure-gauge sector, though further study is required to fully understand the differences in vortex structure between the two sectors that might be responsible for this observation.

Turning our attention to the renormalisation function results shown in Figure~\ref{fig:Z_VR}, we see the respective untouched and vortex-removed renormalisation functions are near identical for $p > 1$~GeV.
The similarity in this region is consistent with previous pure-gauge overlap results as highlighted in Figure~\ref{fig:compareZ}.
We see however, that whilst the untouched and vortex-removed renormalisation functions in the pure-gauge sector are also in agreement for $p<1$~GeV, this is not so in the dynamical case where the renormalisation function is suppressed upon vortex removal.
The novelty of this result, and the decreasing severity of the suppression with increasing quark mass as seen in Figure~\ref{fig:Z_VR} is unsurprising when considered in light of the mass function results.
Since dynamical mass generation vanishes at the lightest masses
\begin{equation}
    M(p) \approx 0 \ \ \forall \, p \,,
    \label{eq:exp2}
\end{equation}
and 
\begin{equation}
    q \to 0 \text{ as } p \to 0 \,,
    \label{eq:exp3}
\end{equation}
it must be that
\begin{equation}
    Z(p) \to 0 \text{ as } p \to 0 \, ,
    \label{eq:exp4}
\end{equation}
to ensure
\begin{equation}
    S(p) = \frac{Z(p)}{i \slashed{q} + M(p)}\,,
    \tag{\ref{eq:S_ctr}}
\end{equation}
remains finite.
In the pure-gauge sector, the imperfect removal of dynamical mass generation leaves $M(0)$ well above zero and therefore removes the restriction of $Z(p) \to 0$ as $p \to 0$. 
Analogously, the reduced suppression at larger quark masses seen in the dynamical results can be seen as a consequence of both the bare quark mass and the greater prominence of remnant dynamical mass generation in the mass function associated with the greater degree of explicit chiral symmetry breaking.

\section{Vortex-only quark propagator}
\label{sec:vortex-only_quark_propagator}

Given the remarkable results of the vortex-removed quark propagator, it is naturally of interest to consider the Landau-gauge overlap quark propagator on a vortex-only background.
However, the guaranteed locality of the overlap operator is predicated on a smoothness condition which is violated by $\Zc(3)$ center-vortex gauge fields. 
It is therefore necessary to smooth them.
The smoothing of $\Zc(3)$ gauge fields is a nontrivial process but viable approaches were found in Ref.~\cite{Virgili:2022ybm}.
In this section we compute the quark propagator on four respective vortex-only ensembles, each smoothed by one of these approaches, in addition to three-loop $\mathcal{O}(a^4)$-improved cooling~\cite{Bilson-Thompson:2002xlt} which was used in the pure-gauge studies of Refs.~\cite{Trewartha:2015ida,Trewartha:2015nna}.
This will serve not only as an initial investigation into the vortex-only quark propagator in dynamical QCD, but may provide further insight to any benefits or drawbacks which may exist in each of the smoothing approaches.

\subsection{Simulation parameters}
\label{subsec:qpvo_sim_params}

We obtain a $\Zc(3)$ vortex-only gauge field ensemble from the same $32^3 \times 64$ PACS-CS 2+1-flavour ensemble~\cite{PACS-CS:2008bkb} used in Ref.~\cite{Virgili:2022wfx} and the previous section ($m_\pi=156$~MeV, $a=0.0933$~fm, $r_0=0.49$~fm).
This vortex-only ensemble is smoothed by each of the respective algorithms developed in Ref.~\cite{Virgili:2022ybm}, and with $\mathcal{O}(a^4)$-improved cooling, such that we obtain four distinct smoothed vortex-only ensembles.
We denote the algorithms as follows:
\begin{description}
\item[CL] $\mathcal{O}(a^4)$-improved cooling,
\item[AS] random-gauge-transformed, APE-style annealed smoothing with smearing parameter $\alpha=0.7$,
\item[CP] as for AS except centrifuge preconditioning is applied prior to the random gauge transformation and smearing parameter $\alpha=0.02$,
\item[VP] as for CP but with vortex-preservation step applied.
\end{description}
These are outlined in Table~\ref{t:qp_recipes}. See Ref.~\cite{Virgili:2022ybm} for details.

In contrast to the untouched and vortex-removed calculations of Ref.~\cite{Virgili:2022wfx} and Section~\ref{sec:vortex-removed}, respectively, which used $H=\gamma^5D_\text{flic}$, here we choose $H~=~\gamma^5D_\text{clover}$.
As the gauge field ensembles have already been smoothed, the additional smoothing in the kernel provided by $D_\text{flic}$ offers no benefit.
For the same reason we also elect to not used mean-field improvement as was done so in the untouched and vortex-removed cases, as the mean link of the smoothed vortex-only ensembles $u_0 \simeq 1$.
For optimal computational efficiency, we choose to project out the 150 lowest-lying eigenmodes, and set the Wilson mass parameter to the canonical value $am_\text{w}=-1.0$, corresponding to a hopping parameter of $\kappa=0.16667$ in the kernel and quark masses $m_q = 5,\,8,\,17,\,25,\,51,\,76$~MeV.

The matrix sign function is calculated using the Zolotarev rational polynomial approximation, and a cylinder cut is applied to the propagator data.

\begin{table}[!b]
    \caption{Summary of smoothing recipes. Steps are applied from left to right starting with the $\Zc(3)$ center-vortex configuration in MCG. C indicates centrifuge preconditioning with rotation angle $\omega$. R indicates the application of a random gauge transformation. $N$ indicates the number of sweeps of cooling (CL) or AUS at smearing parameter $\alpha$. V indicates if a vortex-preservation step was included in the AUS smearing.}
        \label{t:qp_recipes}
        \centering
        \begin{tabular}{ l c c c c c c }
            \noalign{\smallskip}
            \hline\hline
            \noalign{\smallskip}
             Algorithm & C & $\omega$ & R & $N$ & $\alpha$ & V  \\
            \noalign{\smallskip}
            \hline
            \noalign{\smallskip}
             CL & $\times$ & - & $\times$ & 10 & - & $\times$  \\
             AS & $\times$ & - & $\checkmark$ & 20 & 0.7 & $\times$  \\
             CP & $\checkmark$ & 0.02 & $\checkmark$ & 1190 & 0.02 & $\times$  \\
             VP & $\checkmark$ & 0.02 & $\checkmark$ & 1190 & 0.02 & $\checkmark$  \\
            \noalign{\smallskip}
            \hline\hline
            \noalign{\smallskip}
        \end{tabular}
\end{table}

\subsection{Results}

\begin{figure}[!ht]
    \centering
    \includegraphics[width=0.9\linewidth]{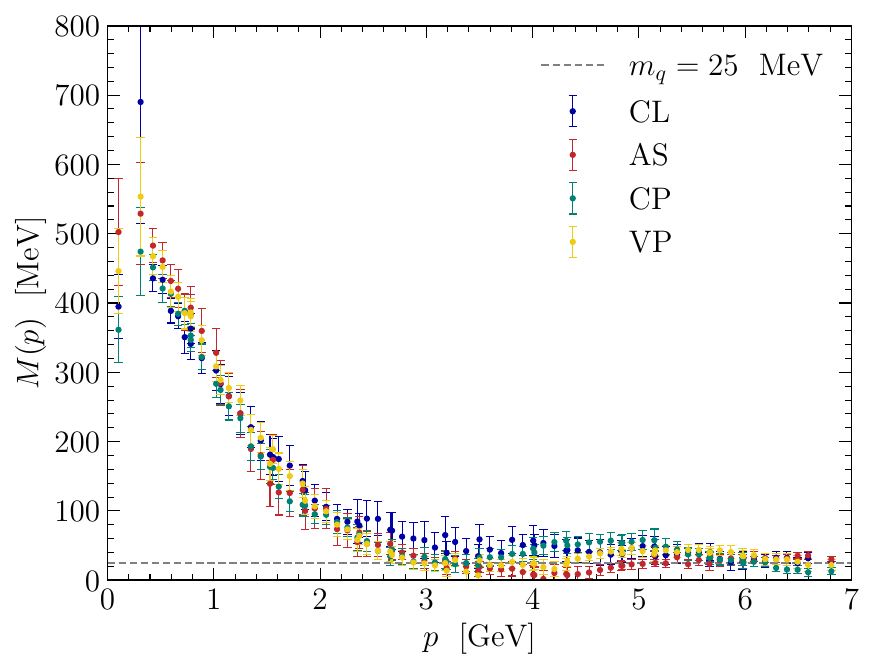}
    \caption{Mass function $M(p)$ for bare quark mass $m_q=25$ MeV. The labels for each simulation are defined in Table~\ref{t:qp_recipes}.}
    \label{fig:M_VO}
\end{figure}
\begin{figure}[!ht]
    \centering
    \includegraphics[width=0.9\linewidth]{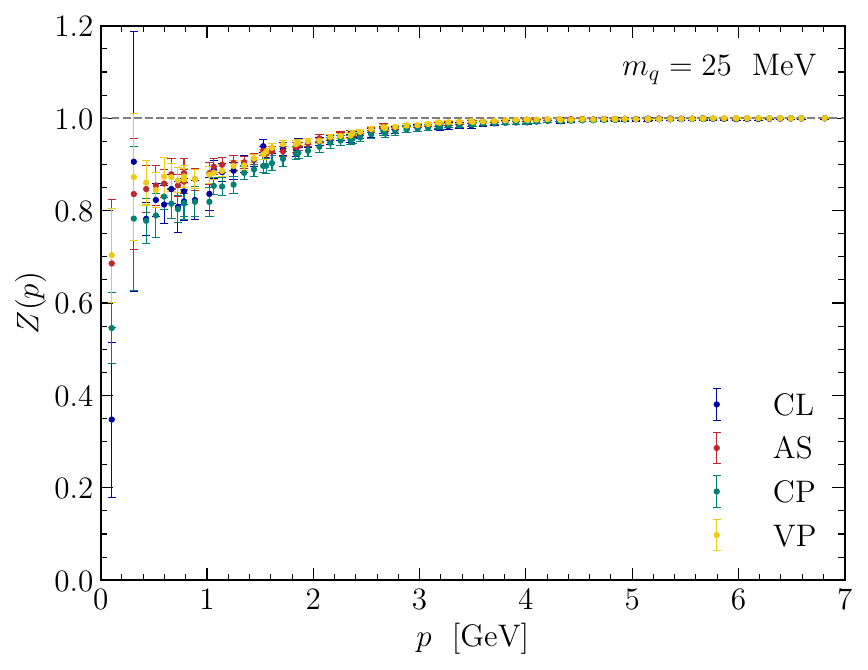}
    \caption{Renormalisation functions$Z(p)$ for for bare quark mass $m_q=25$ MeV. The labels for each simulation are defined in Table~\ref{t:qp_recipes}.}
    \label{fig:Z_VO}
\end{figure}

\begin{figure}[!ht]
    \centering
    \includegraphics[width=0.9\linewidth]{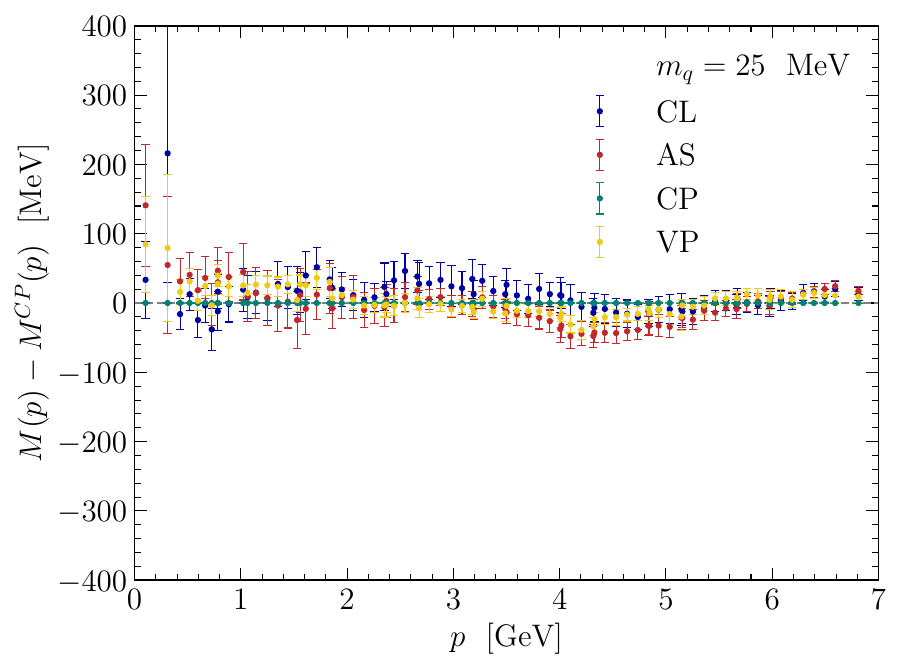}
    \caption{The correlated differences between the respective mass function $M(p)$ from each smoothing algorithm and the CP-smoothed mass function $M^\text{CP}(p)$.}
    \label{fig:MDiff}
\end{figure}
\begin{figure}[!ht]
    \centering
    \includegraphics[width=0.9\linewidth]{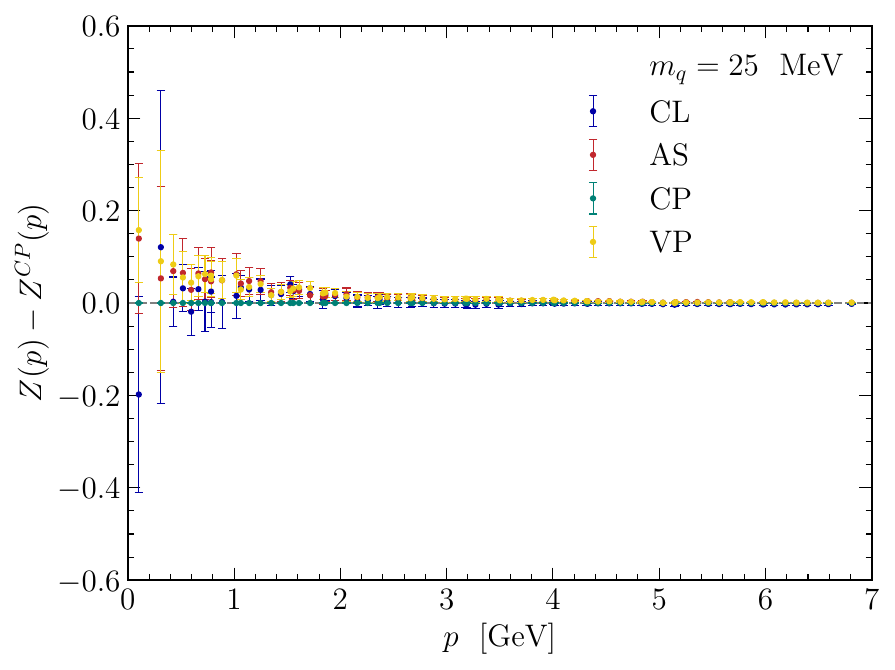}
    \caption{The correlated differences between the respective renormalisation function $Z(p)$ from each smoothing algorithm and the CP-smoothed renormalisation function $Z^\text{CP}(p)$.}
    \label{fig:ZDiff}
\end{figure}

The vortex-only results are significantly noisier than the untouched results of Ref.~\cite{Virgili:2022wfx} despite having the same number of gauge field configurations in the respective ensembles.
Consequently, at the lightest masses any potential differences between the quark propagators derived from each smoothing algorithm are lost in the statistical noise. For this reason, and the sake of brevity, we choose to present only the results and analysis for $m_q=25$~MeV.

The mass and renormalisation functions are plotted in Figs.~\ref{fig:M_VO}~and~\ref{fig:Z_VO}, respectively.
At this mass there are subtle differences between the quark propagators obtained with each smoothing algorithm.
In the mass function, these differences occur around the peak and in the $p=2$--$5$~GeV region, whilst in the renormalisation function they occur in the infrared region.
In an attempt to tease out these differences we consider the correlated differences of the mass and renormalisation functions of each ensemble. 
We choose the CP ensemble as a consistent baseline since this algorithm most closely approaches the ideal of the Wilson flow.

The correlated differences for the mass and renormalisation functions are presented in Figures~\ref{fig:MDiff}~and~\ref{fig:ZDiff}, respectively.
The CL renormalisation function is consistent with CP outside a slight deviation in the $p=1$--$2$~GeV region where it sits high.
Both AS and VP also sit high relative to CP in this region. 
However, whereas CL shows a downward trend back towards CP for $p<1$~GeV, AS and VP appear to continue to diverge on an upward trend away from CP as $p\to0$.
The correlated differences in the mass function reaffirm the subtlety of the differences between the respective ensembles.
The most prominent feature is in the $p=4$--$5$~GeV region where both AS and VP sit low relative to CP.
Outside the neighbourhood of this region, AS is otherwise consistent with CP.
VP, on the other hand, also sits above CP in the $p=1$--$2$~GeV region.
Similarly, CL sits high in the $p=2.5$--$3.5$~GeV region.
This analysis was performed for all masses but did not provide any further insight.
The most notable aspect of these results, however, applies to all ensembles considered. 
The peak in the mass function at small $p$ is much higher than expected.
In this sense, the results of the respective ensembles are broadly similar.

\section{Profile Fitting}
\label{sec:profile_fitting}

To gain further insight into the different smoothed
configurations, it is interesting to explore the nature of the topological objects present.
Here we adopt a similar approach to that taken in Ref.~\cite{Moran:2008qd}, searching the lattice
for sites which are local maxima of the action in their surrounding hypercube. These are then taken
as the approximate centre of an (anti-)instanton-like object, around which we fit the classical
instanton action density
\begin{equation}
S(x) = \xi_S\,\frac{6}{\pi^{2}}\,\frac{\rho_S^{4}}{((x-x_{S})^{2}+\rho_S^{2})^{4}} \, ,
\label{eq:ActionDensity}
\end{equation}
where $\xi_S$, $\rho_S$ and $x_{S}$ are fit parameters.
The parameter $x_{S}$ describes the position of the centre of the action-density profile within the
four-dimensional hypercube such that it is not restricted to a lattice site.
The parameter $\xi_S$ is introduced as the topological objects observed within the lattice
configurations are expected to have a higher action than the classical instanton with $\xi_S =
1$. We wish to characterise the breadth of the object, $\rho_S$, using the shape of the action
density around the local maximum, rather than the height.

With the breadth of the topological object determined by the action density profile, we can then
compare this result with the topological charge density at the origin of the object. For a
classical instanton of charge $Q=\pm 1$ the established relation is
\begin{equation}
q(0)=Q\, \frac{6}{\pi^{2}}\,\frac{1}{\rho_S^{4}} \, .
\label{eq:topcharge}
\end{equation}
This classical relation is plotted in the following figures as black
curves for $Q=\pm1$.

To determine the value of $q(0)$ from the topological charge density of our configurations, two
approaches are considered.  Both approaches commence by fitting a relation analogous to
Eq.~(\ref{eq:ActionDensity}) for the topological charge density
\begin{equation}
q(x) = \xi_q\,\frac{6}{\pi^{2}}\,\frac{\rho_q^{4}}{((x-x_{q})^{2}+\rho_q^{2})^{4}} \, ,
\label{eq:ChargeDensity}
\end{equation}
to the topological charge density calculated on the lattice.  The idea is to provide a functional
form that can interpolate the topological charge density to arbitrary values of $x$.

\begin{figure*}[t]
    \includegraphics[width=0.9\linewidth]{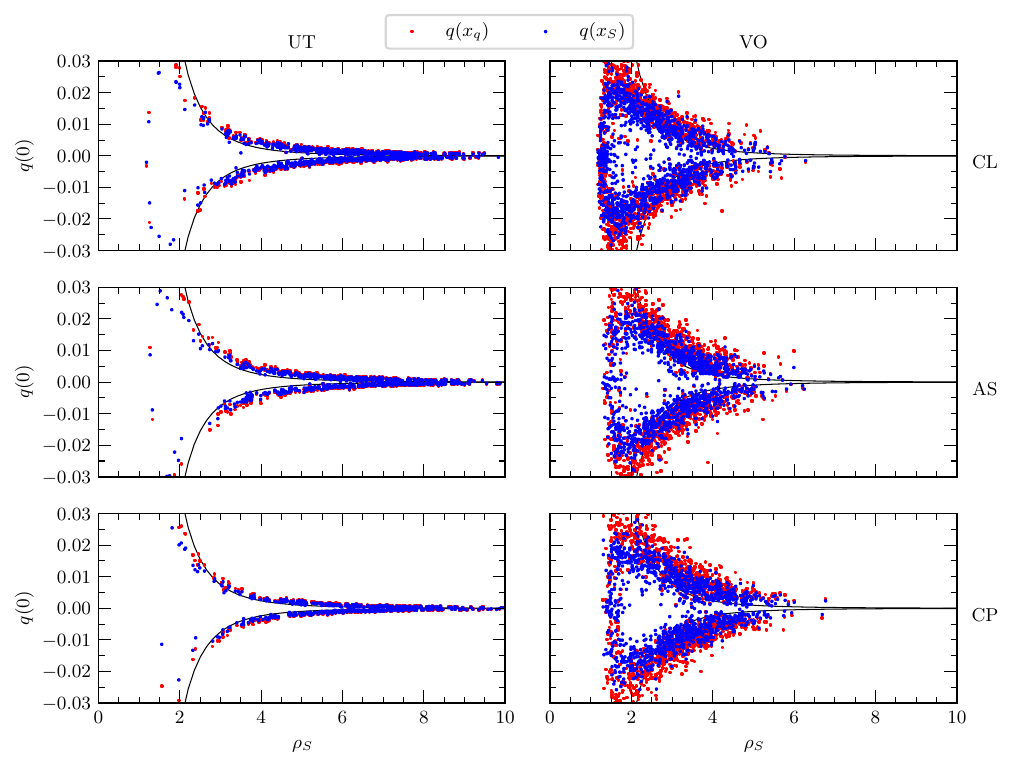}
    \caption{Comparison of $q(0)$ from untouched (left) and vortex-only (right) gauge fields which have been equivalently smoothed using CL (top), AS (middle) and CP (bottom) approaches. Both $q(x_q)$ (red, square) and $q(x_S)$ (blue, circle) are presented. The classical relation is plotted in black. }
    \label{fig:qvrho}
\end{figure*}

\begin{figure}[t]
    \includegraphics[width=0.9\linewidth]{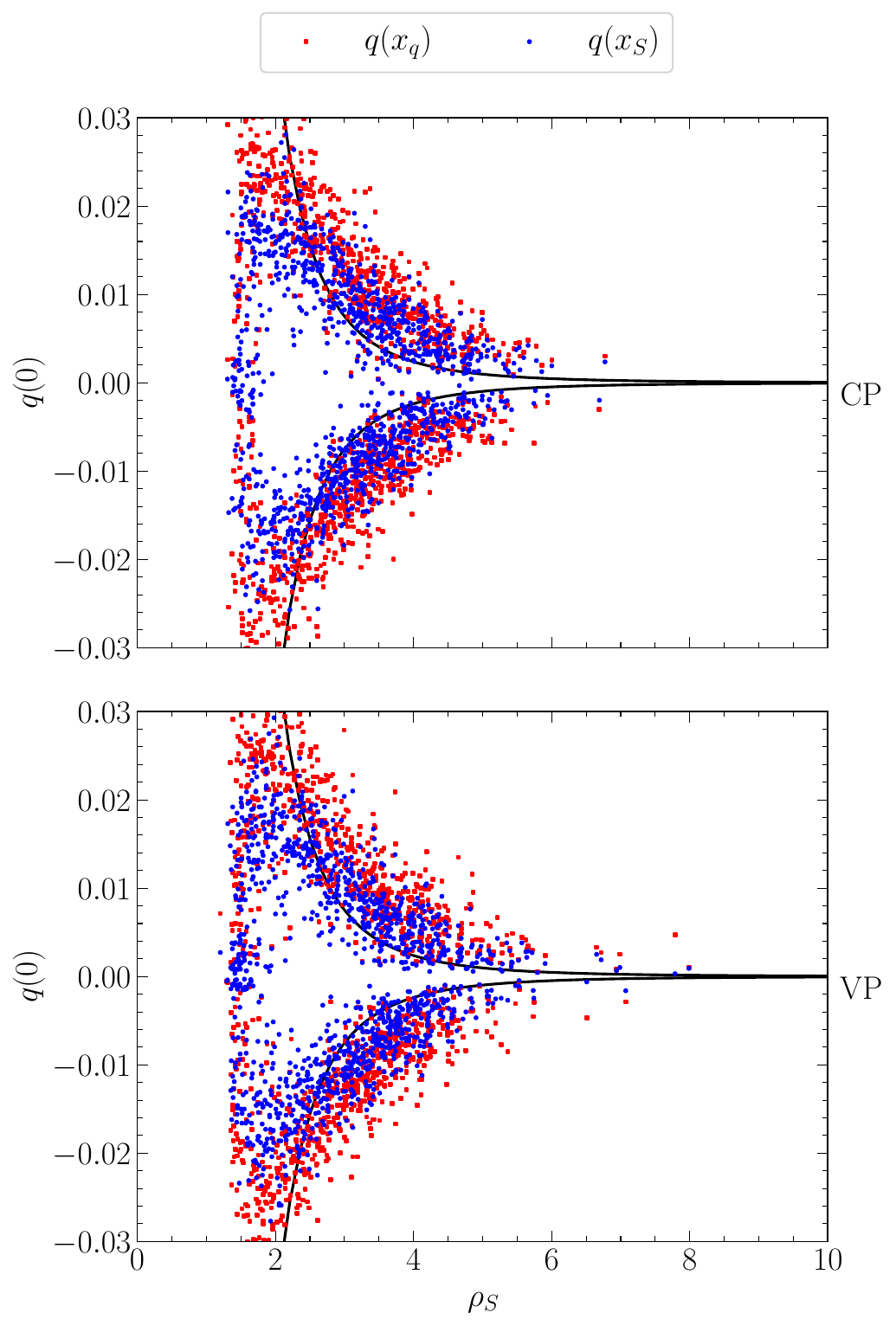}
    \caption{Comparison of $q(0)$ from a vortex-only gauge field which has been smoothed using CP (top) and VP (bottom). Both $q(x_q)$ (red, square) and $q(x_S)$ (blue, circle) are presented. The classical relation is plotted in black. }
    \label{fig:qvrho_VP}
\end{figure}

There are two values of $x$ for $q(x)$ that are of interest.  The first value of interest is
$q(x_{S})$ which estimates the value of the topological charge density at the centre of the
action-density profile.  However, one might wish to allow the peaks in the action and
topological-charge densities to differ slightly and instead report the peak value of the
topological charge density, $q(x_q)$, found within a small hypercube surrounding the local
maximum of the action density.
Both of these methods are used in Figs.~\ref{fig:qvrho} and~\ref{fig:qvrho_VP} to
plot $q$ versus $\rho_S$.

In searching for maxima, we consider each point of the lattice at the centre of a $5^4$ hypercube
and report it as a local maximum if the value exceeds all others within the hypercube.  While it
would be sufficient to consider a $3^4$ hypercube, we find the use of the larger hypercube to be
beneficial in finding maxima in the action density that have a local maxima in the absolute value
of the topological charge density within the same hypercube.  

For the CP algorithm, approximately 3\% of the action
density peaks do not have a corresponding topological charge extrema within the hypercube and these
maxima are discarded from further analysis. This fraction was realised for both the original
untouched gauge fields and the vortex-only fields.  The VP algorithm preserves the roughness of
the vortex-only configurations, such that 6\% of the actions peaks do not have a corresponding
extrema in the topological charge density within the $5^4$ hypercube.  The cooling algorithm sits
between these two with 4-5\% of action peaks not associated with a topological charge density
extrema.

In comparing $\rho_S$ and $\rho_q$, we find $\rho_q \alt \rho_S$ in most cases, and attribute the
smaller value of $\rho_q$ to the need for the topological charge to change sign.  For tightly
packed objects of alternating sign, this can be a significant effect.  Thus we refer to
Eq.~(\ref{eq:topcharge}) as a preferred way to examine the extent to which the topological objects
observed in the lattice gauge fields resemble instantons as depicted by the relation between the
action density size, $\rho_S$, and the topological charge density at the centre of the object.  In
the following we consider both $q(x_{S})$ and $q(x_q)$, where $x_S$ and $x_q$ denote the positions
of the action and topological-charge density centres respectively.


\section{Vortex-Only and Untouched Comparison}
\label{sec:vortex-only_comparison}

We choose the CP smoothing algorithm to compute the vortex-only quark propagator with improved statistics.
Given there are no striking differences between the quark propagators computed in Section~\ref{sec:vortex-only_quark_propagator} which provide a compelling reason to choose one algorithm over another, we choose to employ the CP algorithm as it most closely resembles the ideal of the Wilson flow.

\subsection{Simulation parameters}

We compute the vortex-only quark propagator on a CP-smoothed vortex-only background as described in Subsection~\ref{subsec:qpvo_sim_params} with 2 key differences.
First, we extend the ensemble from the original 30 configurations up to a total of 60, and second, here we choose $am_\text{w}=-1.1$, corresponding to hopping parameter $\kappa=0.17241$ in the kernel.
The former is necessary to reduce the uncertainty in the results to an acceptable level, whilst the latter ensures quark masses match those of of the untouched propagator of Ref.~\cite{Virgili:2022wfx}, which are $m_q = 6,\,9,\,19,\,28,\,56,\,84$ MeV.

\subsection{Results}

\begin{figure*}[!th]
    \centering
    \includegraphics[width=0.9\linewidth]{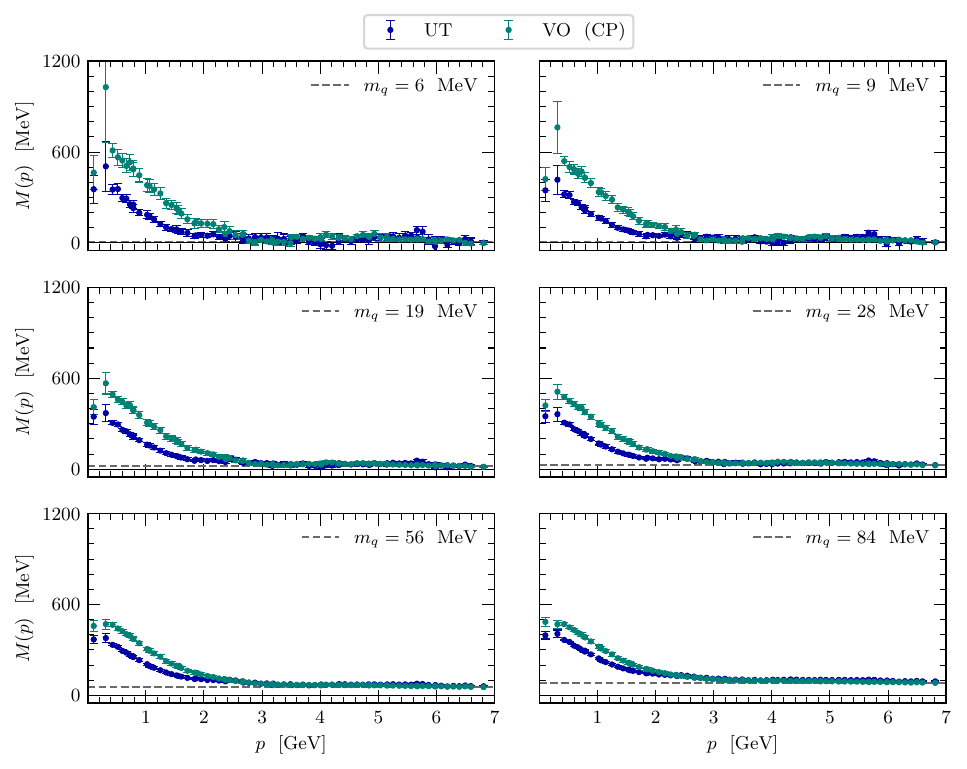}
    \caption{The CP-smoothed vortex-only (green) and untouched (blue) mass functions $M(p)$ for all quark masses considered.}
    \label{fig:Mn60}
\end{figure*}

\begin{figure*}[!th]
    \centering
    \includegraphics[width=0.9\linewidth]{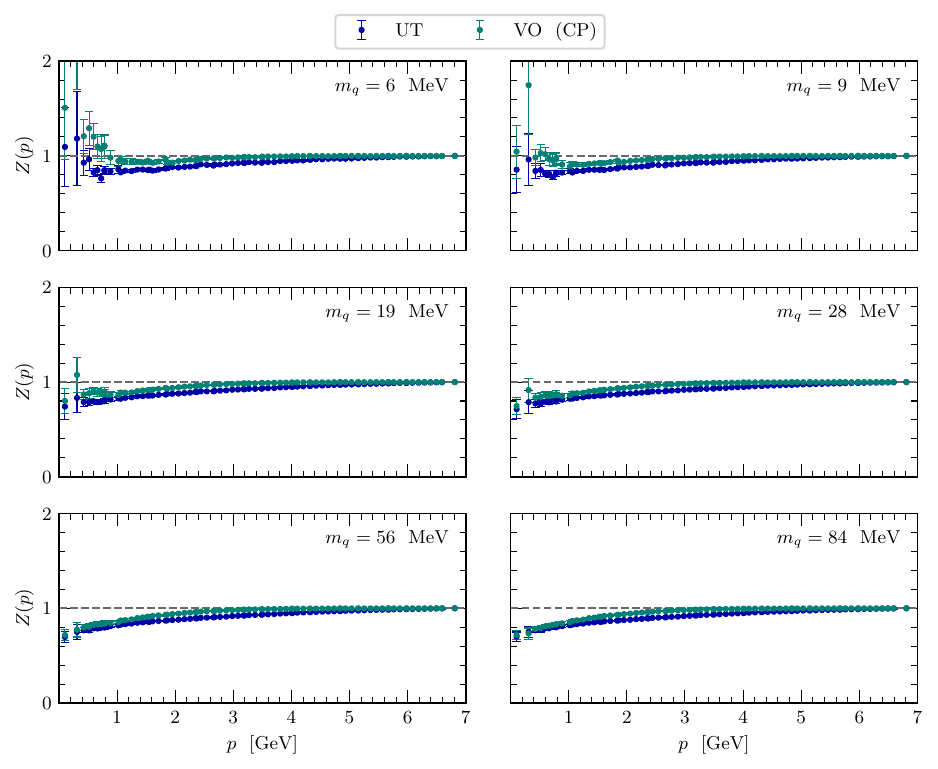}
    \caption{The CP-smoothed vortex-only (green) and untouched (blue) renormalisation functions $Z(p)$ for all quark masses considered.}
    \label{fig:Zn60}
\end{figure*}

The CP-smoothed vortex-only mass and renormalisation functions for each quark mass are plotted against the respective untouched counterparts of Ref.~\cite{Virgili:2022wfx} in Figures~\ref{fig:Mn60}~and~\ref{fig:Zn60}, respectively.

It is in this direct comparison that the apparent excess dynamical mass generation in the vortex-only mass function becomes clear.
Even against the unsmoothed untouched mass function, the vortex-only mass function sits consistently higher in the region in which dynamical mass generation occurs.
This is in contrast to the pure-gauge sector where the dynamical mass generation in the vortex-only mass function matches that of the equivalently-smoothed untouched mass function~\cite{Trewartha:2015nna}.

There is a clear peak in the vortex-only mass function at lighter masses, which becomes less prominent with increasing quark mass, ultimately reaching a plateau at the heaviest quark mass.
Whilst the height of the peak is mass dependent, the data point at the smallest momentum is independent of valence quark mass. 

The relative degree of `excess' dynamical mass generation in the $0.5 < p < 2.5$ increases with decreasing quark mass.
The separation between the respective mass functions shifts rightward with decreasing mass, separating at $p \sim 2$~GeV at the heaviest mass, but closer to $p \sim 2.5$~GeV at the lightest.
This is of potential interest given that the quark mass serves as a measure of explicit chiral symmetry breaking.

Meanwhile, the direct comparison of the respective vortex-only and untouched renormalisation functions reveals a divergence between the two.
The degree of divergence appears consistent across the different quark masses in the $p=1.5$--$5$~GeV range.
At all but the heaviest masses, this divergence persists into the infrared with the vortex-only function sitting consistently above the untouched.
As such, like the untouched, there is an uptick in the vortex-only function at the lightest masses.
However, the minimum prior to the uptick is shifted rightward relative to the untouched, and is less prominent making it more difficult to precisely pinpoint.
Like the untouched case, the uptick vanishes at heavier masses, to the point that the respective functions are in agreement in the region at the heaviest mass.
For $p>5$~GeV the respective functions are in agreement.


\section{Conclusions}
\label{sec:conclusions}

The majority of these studies were conducted in the pure-gauge sector, with some consequential discrepancies in their results. Recent studies of the static quark potential, gluon propagator, and quark mass function found the presence of dynamical quarks resolved many of the discrepancies that appear in the pure-gauge sector. In particular, in the pure gauge sector a remnant of dynamical mass generation persists in the quark mass function after vortex removal.

In the pure-gauge sector, a significant remnant of dynamical mass generation in the quark propagator mass function persists upon vortex-removal.
This is despite the full reproduction of dynamical mass generation on a vortex-only background.
Recent studies of the static quark potential and gluon propagator found that the presence of dynamical fermions resolves similar quantitative discrepancies related to vortex modification in the pure-gauge sector.
Motivated by these results, it was investigated whether dynamical fermions could play a similar role in the context of the vortex-modified quark propagator. 

To this end, the Landau-gauge overlap quark propagator was computed on a vortex-removed 2+1 dynamical fermion flavour gauge field ensemble.
Dynamical mass generation in the mass function vanishes upon vortex removal at the lightest quark mass, demonstrating an absence of dynamical chiral symmetry breaking and resolving the discrepancy present in the pure-gauge sector.
Furthermore the quark mass governs the degree to which chiral symmetry is explicitly broken. 
As the quark mass increases, a remnant of dynamical mass generation emerges and becomes more prominent, but remains small.
Within this framework, it is clear that the novel infrared suppression in the renormalisation function upon vortex-removal is necessary for the quark propagator to remain finite.

The Landau-gauge overlap quark propagator was also computed on smoothed vortex-only ensembles obtained from 2+1-flavour dynamical gauge fields.
Respective ensembles were smoothed by each of the recipes developed in Ref.~\cite{Virgili:2022ybm}, in addition to $\mathcal{O}(a^4)$-improved cooling.
For direct comparison with the untouched propagator of Ref.~\cite{Virgili:2022wfx}, a more precise calculation was performed using the CP smoothing recipe with an additional 30 gauge field configurations for an ensemble total of 60.

Dynamical mass generation in the mass function is qualitatively reproduced on the vortex-only background, consistent with the pure-gauge sector.
Unlike the pure-gauge sector, where the amount of dynamical mass generation on the vortex-only background is consistent with that of an equivalently smoothed untouched background, excess dynamical mass generation is observed in the vortex-only mass function -- even when compared to the unsmoothed untouched mass function.
The relative degree of the excess dynamical mass generation appears to have a quark mass dependence, being larger at lighter masses.
These results appear to be independent of choice of smoothing algorithm.
It is difficult to tease out any subtle differences which may be present between the various smoothing algorithms.

The presence of dynamical fermions resolves the quantitative discrepancy observed in the pure-gauge vortex-removed quark propagator.
The cause and mechanism of the excess dynamical mass generation from the vortex-only background is unknown, and is a topic of further research.
Together, these results support an important relationship between dynamical fermions and centre vortices, and further add to the body of evidence supporting centre vortices as the mechanism underpinning dynamical chiral symmetry breaking.

The \textsc{cola} software library~\cite{Kamleh:2022nqr} was used to calculate the lattice results reported herein.

\begin{acknowledgments}
We thank the PACS-CS Collaboration for making their configurations available via the International Lattice Data Grid (ILDG).
This research was undertaken with resources provided by the Pawsey Supercomputing Centre through the National Computational Merit Allocation Scheme with funding from the Australian Government and the Government of Western Australia. Additional resources were provided from the National Computational Infrastructure (NCI) supported by the Australian Government through Grant No. LE190100021 via the University of Adelaide Partner Share.
This research is supported by Australian Research Council through Grants No. DP190102215 and DP210103706.
WK is supported by the Pawsey Supercomputing Centre through the Pawsey Centre for Extreme Scale Readiness (PaCER) program.
\end{acknowledgments}


\begin{thebibliography}{72}%
\makeatletter
\providecommand \@ifxundefined [1]{%
 \@ifx{#1\undefined}
}%
\providecommand \@ifnum [1]{%
 \ifnum #1\expandafter \@firstoftwo
 \else \expandafter \@secondoftwo
 \fi
}%
\providecommand \@ifx [1]{%
 \ifx #1\expandafter \@firstoftwo
 \else \expandafter \@secondoftwo
 \fi
}%
\providecommand \natexlab [1]{#1}%
\providecommand \enquote  [1]{``#1''}%
\providecommand \bibnamefont  [1]{#1}%
\providecommand \bibfnamefont [1]{#1}%
\providecommand \citenamefont [1]{#1}%
\providecommand \href@noop [0]{\@secondoftwo}%
\providecommand \href [0]{\begingroup \@sanitize@url \@href}%
\providecommand \@href[1]{\@@startlink{#1}\@@href}%
\providecommand \@@href[1]{\endgroup#1\@@endlink}%
\providecommand \@sanitize@url [0]{\catcode `\\12\catcode `\$12\catcode
  `\&12\catcode `\#12\catcode `\^12\catcode `\_12\catcode `\%12\relax}%
\providecommand \@@startlink[1]{}%
\providecommand \@@endlink[0]{}%
\providecommand \url  [0]{\begingroup\@sanitize@url \@url }%
\providecommand \@url [1]{\endgroup\@href {#1}{\urlprefix }}%
\providecommand \urlprefix  [0]{URL }%
\providecommand \Eprint [0]{\href }%
\providecommand \doibase [0]{http://dx.doi.org/}%
\providecommand \selectlanguage [0]{\@gobble}%
\providecommand \bibinfo  [0]{\@secondoftwo}%
\providecommand \bibfield  [0]{\@secondoftwo}%
\providecommand \translation [1]{[#1]}%
\providecommand \BibitemOpen [0]{}%
\providecommand \bibitemStop [0]{}%
\providecommand \bibitemNoStop [0]{.\EOS\space}%
\providecommand \EOS [0]{\spacefactor3000\relax}%
\providecommand \BibitemShut  [1]{\csname bibitem#1\endcsname}%
\let\auto@bib@innerbib\@empty
\bibitem [{\citenamefont {Kamleh}\ \emph {et~al.}(2024)\citenamefont {Kamleh},
  \citenamefont {Leinweber},\ and\ \citenamefont {Virgili}}]{Kamleh:2023gho}%
  \BibitemOpen
  \bibfield  {author} {\bibinfo {author} {\bibfnamefont {W.}~\bibnamefont
  {Kamleh}}, \bibinfo {author} {\bibfnamefont {D.~B.}\ \bibnamefont
  {Leinweber}}, \ and\ \bibinfo {author} {\bibfnamefont {A.}~\bibnamefont
  {Virgili}},\ }\href {\doibase 10.1103/PhysRevD.110.L051502} {\bibfield
  {journal} {\bibinfo  {journal} {Phys. Rev. D}\ }\textbf {\bibinfo {volume}
  {110}},\ \bibinfo {pages} {L051502} (\bibinfo {year} {2024})},\ \Eprint
  {http://arxiv.org/abs/2305.18690} {arXiv:2305.18690 [hep-lat]} \BibitemShut
  {NoStop}%
\bibitem [{\citenamefont {'t~Hooft}(1978)}]{tHooft:1977nqb}%
  \BibitemOpen
  \bibfield  {author} {\bibinfo {author} {\bibfnamefont {G.}~\bibnamefont
  {'t~Hooft}},\ }\href {\doibase 10.1016/0550-3213(78)90153-0} {\bibfield
  {journal} {\bibinfo  {journal} {Nucl. Phys. B}\ }\textbf {\bibinfo {volume}
  {138}},\ \bibinfo {pages} {1} (\bibinfo {year} {1978})}\BibitemShut {NoStop}%
\bibitem [{\citenamefont {'t~Hooft}(1979)}]{tHooft:1979rtg}%
  \BibitemOpen
  \bibfield  {author} {\bibinfo {author} {\bibfnamefont {G.}~\bibnamefont
  {'t~Hooft}},\ }\href {\doibase 10.1016/0550-3213(79)90595-9} {\bibfield
  {journal} {\bibinfo  {journal} {Nucl. Phys. B}\ }\textbf {\bibinfo {volume}
  {153}},\ \bibinfo {pages} {141} (\bibinfo {year} {1979})}\BibitemShut
  {NoStop}%
\bibitem [{\citenamefont {Del~Debbio}\ \emph {et~al.}(1997)\citenamefont
  {Del~Debbio}, \citenamefont {Faber}, \citenamefont {Greensite},\ and\
  \citenamefont {Olejnik}}]{DelDebbio:1996lih}%
  \BibitemOpen
  \bibfield  {author} {\bibinfo {author} {\bibfnamefont {L.}~\bibnamefont
  {Del~Debbio}}, \bibinfo {author} {\bibfnamefont {M.}~\bibnamefont {Faber}},
  \bibinfo {author} {\bibfnamefont {J.}~\bibnamefont {Greensite}}, \ and\
  \bibinfo {author} {\bibfnamefont {S.}~\bibnamefont {Olejnik}},\ }\href
  {\doibase 10.1103/PhysRevD.55.2298} {\bibfield  {journal} {\bibinfo
  {journal} {Phys. Rev. D}\ }\textbf {\bibinfo {volume} {55}},\ \bibinfo
  {pages} {2298} (\bibinfo {year} {1997})},\ \Eprint
  {http://arxiv.org/abs/hep-lat/9610005} {arXiv:hep-lat/9610005} \BibitemShut
  {NoStop}%
\bibitem [{\citenamefont {Faber}\ \emph {et~al.}(1998)\citenamefont {Faber},
  \citenamefont {Greensite},\ and\ \citenamefont {Olejnik}}]{Faber:1997rp}%
  \BibitemOpen
  \bibfield  {author} {\bibinfo {author} {\bibfnamefont {M.}~\bibnamefont
  {Faber}}, \bibinfo {author} {\bibfnamefont {J.}~\bibnamefont {Greensite}}, \
  and\ \bibinfo {author} {\bibfnamefont {S.}~\bibnamefont {Olejnik}},\ }\href
  {\doibase 10.1103/PhysRevD.57.2603} {\bibfield  {journal} {\bibinfo
  {journal} {Phys. Rev. D}\ }\textbf {\bibinfo {volume} {57}},\ \bibinfo
  {pages} {2603} (\bibinfo {year} {1998})},\ \Eprint
  {http://arxiv.org/abs/hep-lat/9710039} {arXiv:hep-lat/9710039} \BibitemShut
  {NoStop}%
\bibitem [{\citenamefont {Del~Debbio}\ \emph {et~al.}(1998)\citenamefont
  {Del~Debbio}, \citenamefont {Faber}, \citenamefont {Giedt}, \citenamefont
  {Greensite},\ and\ \citenamefont {Olejnik}}]{DelDebbio:1998luz}%
  \BibitemOpen
  \bibfield  {author} {\bibinfo {author} {\bibfnamefont {L.}~\bibnamefont
  {Del~Debbio}}, \bibinfo {author} {\bibfnamefont {M.}~\bibnamefont {Faber}},
  \bibinfo {author} {\bibfnamefont {J.}~\bibnamefont {Giedt}}, \bibinfo
  {author} {\bibfnamefont {J.}~\bibnamefont {Greensite}}, \ and\ \bibinfo
  {author} {\bibfnamefont {S.}~\bibnamefont {Olejnik}},\ }\href {\doibase
  10.1103/PhysRevD.58.094501} {\bibfield  {journal} {\bibinfo  {journal} {Phys.
  Rev. D}\ }\textbf {\bibinfo {volume} {58}},\ \bibinfo {pages} {094501}
  (\bibinfo {year} {1998})},\ \Eprint {http://arxiv.org/abs/hep-lat/9801027}
  {arXiv:hep-lat/9801027} \BibitemShut {NoStop}%
\bibitem [{\citenamefont {Bertle}\ \emph {et~al.}(1999)\citenamefont {Bertle},
  \citenamefont {Faber}, \citenamefont {Greensite},\ and\ \citenamefont
  {Olejnik}}]{Bertle:1999tw}%
  \BibitemOpen
  \bibfield  {author} {\bibinfo {author} {\bibfnamefont {R.}~\bibnamefont
  {Bertle}}, \bibinfo {author} {\bibfnamefont {M.}~\bibnamefont {Faber}},
  \bibinfo {author} {\bibfnamefont {J.}~\bibnamefont {Greensite}}, \ and\
  \bibinfo {author} {\bibfnamefont {S.}~\bibnamefont {Olejnik}},\ }\href
  {\doibase 10.1088/1126-6708/1999/03/019} {\bibfield  {journal} {\bibinfo
  {journal} {JHEP}\ }\textbf {\bibinfo {volume} {03}},\ \bibinfo {pages} {019}
  (\bibinfo {year} {1999})},\ \Eprint {http://arxiv.org/abs/hep-lat/9903023}
  {arXiv:hep-lat/9903023} \BibitemShut {NoStop}%
\bibitem [{\citenamefont {de~Forcrand}\ and\ \citenamefont
  {D'Elia}(1999)}]{deForcrand:1999our}%
  \BibitemOpen
  \bibfield  {author} {\bibinfo {author} {\bibfnamefont {P.}~\bibnamefont
  {de~Forcrand}}\ and\ \bibinfo {author} {\bibfnamefont {M.}~\bibnamefont
  {D'Elia}},\ }\href {\doibase 10.1103/PhysRevLett.82.4582} {\bibfield
  {journal} {\bibinfo  {journal} {Phys. Rev. Lett.}\ }\textbf {\bibinfo
  {volume} {82}},\ \bibinfo {pages} {4582} (\bibinfo {year} {1999})},\ \Eprint
  {http://arxiv.org/abs/hep-lat/9901020} {arXiv:hep-lat/9901020} \BibitemShut
  {NoStop}%
\bibitem [{\citenamefont {Faber}\ \emph {et~al.}(1999)\citenamefont {Faber},
  \citenamefont {Greensite}, \citenamefont {Olejnik},\ and\ \citenamefont
  {Yamada}}]{Faber:1999gu}%
  \BibitemOpen
  \bibfield  {author} {\bibinfo {author} {\bibfnamefont {M.}~\bibnamefont
  {Faber}}, \bibinfo {author} {\bibfnamefont {J.}~\bibnamefont {Greensite}},
  \bibinfo {author} {\bibfnamefont {S.}~\bibnamefont {Olejnik}}, \ and\
  \bibinfo {author} {\bibfnamefont {D.}~\bibnamefont {Yamada}},\ }\href
  {\doibase 10.1088/1126-6708/1999/12/012} {\bibfield  {journal} {\bibinfo
  {journal} {JHEP}\ }\textbf {\bibinfo {volume} {12}},\ \bibinfo {pages} {012}
  (\bibinfo {year} {1999})},\ \Eprint {http://arxiv.org/abs/hep-lat/9910033}
  {arXiv:hep-lat/9910033} \BibitemShut {NoStop}%
\bibitem [{\citenamefont {Engelhardt}\ \emph {et~al.}(2000)\citenamefont
  {Engelhardt}, \citenamefont {Langfeld}, \citenamefont {Reinhardt},\ and\
  \citenamefont {Tennert}}]{Engelhardt:1999fd}%
  \BibitemOpen
  \bibfield  {author} {\bibinfo {author} {\bibfnamefont {M.}~\bibnamefont
  {Engelhardt}}, \bibinfo {author} {\bibfnamefont {K.}~\bibnamefont
  {Langfeld}}, \bibinfo {author} {\bibfnamefont {H.}~\bibnamefont {Reinhardt}},
  \ and\ \bibinfo {author} {\bibfnamefont {O.}~\bibnamefont {Tennert}},\ }\href
  {\doibase 10.1103/PhysRevD.61.054504} {\bibfield  {journal} {\bibinfo
  {journal} {Phys. Rev. D}\ }\textbf {\bibinfo {volume} {61}},\ \bibinfo
  {pages} {054504} (\bibinfo {year} {2000})},\ \Eprint
  {http://arxiv.org/abs/hep-lat/9904004} {arXiv:hep-lat/9904004} \BibitemShut
  {NoStop}%
\bibitem [{\citenamefont {Engelhardt}\ and\ \citenamefont
  {Reinhardt}(2000)}]{Engelhardt:1999xw}%
  \BibitemOpen
  \bibfield  {author} {\bibinfo {author} {\bibfnamefont {M.}~\bibnamefont
  {Engelhardt}}\ and\ \bibinfo {author} {\bibfnamefont {H.}~\bibnamefont
  {Reinhardt}},\ }\href {\doibase 10.1016/S0550-3213(99)00727-0} {\bibfield
  {journal} {\bibinfo  {journal} {Nucl. Phys. B}\ }\textbf {\bibinfo {volume}
  {567}},\ \bibinfo {pages} {249} (\bibinfo {year} {2000})},\ \Eprint
  {http://arxiv.org/abs/hep-th/9907139} {arXiv:hep-th/9907139} \BibitemShut
  {NoStop}%
\bibitem [{\citenamefont {Engelhardt}(2000)}]{Engelhardt:2000wc}%
  \BibitemOpen
  \bibfield  {author} {\bibinfo {author} {\bibfnamefont {M.}~\bibnamefont
  {Engelhardt}},\ }\href {\doibase 10.1016/S0550-3213(00)00350-3} {\bibfield
  {journal} {\bibinfo  {journal} {Nucl. Phys.}\ }\textbf {\bibinfo {volume}
  {B585}},\ \bibinfo {pages} {614} (\bibinfo {year} {2000})},\ \Eprint
  {http://arxiv.org/abs/hep-lat/0004013} {arXiv:hep-lat/0004013 [hep-lat]}
  \BibitemShut {NoStop}%
\bibitem [{\citenamefont {Bertle}\ \emph {et~al.}(2000)\citenamefont {Bertle},
  \citenamefont {Faber}, \citenamefont {Greensite},\ and\ \citenamefont
  {Olejnik}}]{Bertle:2000qv}%
  \BibitemOpen
  \bibfield  {author} {\bibinfo {author} {\bibfnamefont {R.}~\bibnamefont
  {Bertle}}, \bibinfo {author} {\bibfnamefont {M.}~\bibnamefont {Faber}},
  \bibinfo {author} {\bibfnamefont {J.}~\bibnamefont {Greensite}}, \ and\
  \bibinfo {author} {\bibfnamefont {S.}~\bibnamefont {Olejnik}},\ }\href
  {\doibase 10.1088/1126-6708/2000/10/007} {\bibfield  {journal} {\bibinfo
  {journal} {JHEP}\ }\textbf {\bibinfo {volume} {10}},\ \bibinfo {pages} {007}
  (\bibinfo {year} {2000})},\ \Eprint {http://arxiv.org/abs/hep-lat/0007043}
  {arXiv:hep-lat/0007043} \BibitemShut {NoStop}%
\bibitem [{\citenamefont {Langfeld}\ \emph {et~al.}(2002)\citenamefont
  {Langfeld}, \citenamefont {Reinhardt},\ and\ \citenamefont
  {Gattnar}}]{Langfeld:2001cz}%
  \BibitemOpen
  \bibfield  {author} {\bibinfo {author} {\bibfnamefont {K.}~\bibnamefont
  {Langfeld}}, \bibinfo {author} {\bibfnamefont {H.}~\bibnamefont {Reinhardt}},
  \ and\ \bibinfo {author} {\bibfnamefont {J.}~\bibnamefont {Gattnar}},\ }\href
  {\doibase 10.1016/S0550-3213(01)00574-0} {\bibfield  {journal} {\bibinfo
  {journal} {Nucl. Phys. B}\ }\textbf {\bibinfo {volume} {621}},\ \bibinfo
  {pages} {131} (\bibinfo {year} {2002})},\ \Eprint
  {http://arxiv.org/abs/hep-ph/0107141} {arXiv:hep-ph/0107141} \BibitemShut
  {NoStop}%
\bibitem [{\citenamefont {Engelhardt}(2002)}]{Engelhardt:2002qs}%
  \BibitemOpen
  \bibfield  {author} {\bibinfo {author} {\bibfnamefont {M.}~\bibnamefont
  {Engelhardt}},\ }\href {\doibase 10.1016/S0550-3213(02)00470-4} {\bibfield
  {journal} {\bibinfo  {journal} {Nucl. Phys. B}\ }\textbf {\bibinfo {volume}
  {638}},\ \bibinfo {pages} {81} (\bibinfo {year} {2002})},\ \Eprint
  {http://arxiv.org/abs/hep-lat/0204002} {arXiv:hep-lat/0204002} \BibitemShut
  {NoStop}%
\bibitem [{\citenamefont {Langfeld}(2004)}]{Langfeld:2003ev}%
  \BibitemOpen
  \bibfield  {author} {\bibinfo {author} {\bibfnamefont {K.}~\bibnamefont
  {Langfeld}},\ }\href {\doibase 10.1103/PhysRevD.69.014503} {\bibfield
  {journal} {\bibinfo  {journal} {Phys. Rev. D}\ }\textbf {\bibinfo {volume}
  {69}},\ \bibinfo {pages} {014503} (\bibinfo {year} {2004})},\ \Eprint
  {http://arxiv.org/abs/hep-lat/0307030} {arXiv:hep-lat/0307030} \BibitemShut
  {NoStop}%
\bibitem [{\citenamefont {Greensite}(2003)}]{Greensite:2003bk}%
  \BibitemOpen
  \bibfield  {author} {\bibinfo {author} {\bibfnamefont {J.}~\bibnamefont
  {Greensite}},\ }\href {\doibase 10.1016/S0146-6410(03)90012-3} {\bibfield
  {journal} {\bibinfo  {journal} {Prog. Part. Nucl. Phys.}\ }\textbf {\bibinfo
  {volume} {51}},\ \bibinfo {pages} {1} (\bibinfo {year} {2003})},\ \Eprint
  {http://arxiv.org/abs/hep-lat/0301023} {arXiv:hep-lat/0301023} \BibitemShut
  {NoStop}%
\bibitem [{\citenamefont {Bruckmann}\ and\ \citenamefont
  {Engelhardt}(2003)}]{Bruckmann:2003yd}%
  \BibitemOpen
  \bibfield  {author} {\bibinfo {author} {\bibfnamefont {F.}~\bibnamefont
  {Bruckmann}}\ and\ \bibinfo {author} {\bibfnamefont {M.}~\bibnamefont
  {Engelhardt}},\ }\href {\doibase 10.1103/PhysRevD.68.105011} {\bibfield
  {journal} {\bibinfo  {journal} {Phys. Rev.}\ }\textbf {\bibinfo {volume}
  {D68}},\ \bibinfo {pages} {105011} (\bibinfo {year} {2003})},\ \Eprint
  {http://arxiv.org/abs/hep-th/0307219} {arXiv:hep-th/0307219 [hep-th]}
  \BibitemShut {NoStop}%
\bibitem [{\citenamefont {Engelhardt}\ \emph {et~al.}(2004)\citenamefont
  {Engelhardt}, \citenamefont {Quandt},\ and\ \citenamefont
  {Reinhardt}}]{Engelhardt:2003wm}%
  \BibitemOpen
  \bibfield  {author} {\bibinfo {author} {\bibfnamefont {M.}~\bibnamefont
  {Engelhardt}}, \bibinfo {author} {\bibfnamefont {M.}~\bibnamefont {Quandt}},
  \ and\ \bibinfo {author} {\bibfnamefont {H.}~\bibnamefont {Reinhardt}},\
  }\href {\doibase 10.1016/j.nuclphysb.2004.02.036} {\bibfield  {journal}
  {\bibinfo  {journal} {Nucl. Phys. B}\ }\textbf {\bibinfo {volume} {685}},\
  \bibinfo {pages} {227} (\bibinfo {year} {2004})},\ \Eprint
  {http://arxiv.org/abs/hep-lat/0311029} {arXiv:hep-lat/0311029} \BibitemShut
  {NoStop}%
\bibitem [{\citenamefont {Boyko}\ \emph {et~al.}(2006)\citenamefont {Boyko},
  \citenamefont {Bornyakov}, \citenamefont {Ilgenfritz}, \citenamefont
  {Kovalenko}, \citenamefont {Martemyanov}, \citenamefont {Muller-Preussker},
  \citenamefont {Polikarpov},\ and\ \citenamefont {Veselov}}]{Boyko:2006ic}%
  \BibitemOpen
  \bibfield  {author} {\bibinfo {author} {\bibfnamefont {P.~Y.}\ \bibnamefont
  {Boyko}}, \bibinfo {author} {\bibfnamefont {V.~G.}\ \bibnamefont
  {Bornyakov}}, \bibinfo {author} {\bibfnamefont {E.~M.}\ \bibnamefont
  {Ilgenfritz}}, \bibinfo {author} {\bibfnamefont {A.~V.}\ \bibnamefont
  {Kovalenko}}, \bibinfo {author} {\bibfnamefont {B.~V.}\ \bibnamefont
  {Martemyanov}}, \bibinfo {author} {\bibfnamefont {M.}~\bibnamefont
  {Muller-Preussker}}, \bibinfo {author} {\bibfnamefont {M.~I.}\ \bibnamefont
  {Polikarpov}}, \ and\ \bibinfo {author} {\bibfnamefont {A.~I.}\ \bibnamefont
  {Veselov}},\ }\href {\doibase 10.1016/j.nuclphysb.2006.08.025} {\bibfield
  {journal} {\bibinfo  {journal} {Nucl. Phys. B}\ }\textbf {\bibinfo {volume}
  {756}},\ \bibinfo {pages} {71} (\bibinfo {year} {2006})},\ \Eprint
  {http://arxiv.org/abs/hep-lat/0607003} {arXiv:hep-lat/0607003} \BibitemShut
  {NoStop}%
\bibitem [{\citenamefont {Ilgenfritz}\ \emph {et~al.}(2007)\citenamefont
  {Ilgenfritz}, \citenamefont {Koller}, \citenamefont {Koma}, \citenamefont
  {Schierholz}, \citenamefont {Streuer}, \citenamefont {Weinberg},\ and\
  \citenamefont {Quandt}}]{Ilgenfritz:2007ua}%
  \BibitemOpen
  \bibfield  {author} {\bibinfo {author} {\bibfnamefont {E.-M.}\ \bibnamefont
  {Ilgenfritz}}, \bibinfo {author} {\bibfnamefont {K.}~\bibnamefont {Koller}},
  \bibinfo {author} {\bibfnamefont {Y.}~\bibnamefont {Koma}}, \bibinfo {author}
  {\bibfnamefont {G.}~\bibnamefont {Schierholz}}, \bibinfo {author}
  {\bibfnamefont {T.}~\bibnamefont {Streuer}}, \bibinfo {author} {\bibfnamefont
  {V.}~\bibnamefont {Weinberg}}, \ and\ \bibinfo {author} {\bibfnamefont
  {M.}~\bibnamefont {Quandt}},\ }\href {\doibase 10.22323/1.042.0311}
  {\bibfield  {journal} {\bibinfo  {journal} {PoS}\ }\textbf {\bibinfo {volume}
  {LATTICE2007}},\ \bibinfo {pages} {311} (\bibinfo {year} {2007})},\ \Eprint
  {http://arxiv.org/abs/0710.2607} {arXiv:0710.2607 [hep-lat]} \BibitemShut
  {NoStop}%
\bibitem [{\citenamefont {Bornyakov}\ \emph {et~al.}(2008)\citenamefont
  {Bornyakov}, \citenamefont {Ilgenfritz}, \citenamefont {Martemyanov},
  \citenamefont {Morozov}, \citenamefont {Muller-Preussker},\ and\
  \citenamefont {Veselov}}]{Bornyakov:2007fz}%
  \BibitemOpen
  \bibfield  {author} {\bibinfo {author} {\bibfnamefont {V.~G.}\ \bibnamefont
  {Bornyakov}}, \bibinfo {author} {\bibfnamefont {E.~M.}\ \bibnamefont
  {Ilgenfritz}}, \bibinfo {author} {\bibfnamefont {B.~V.}\ \bibnamefont
  {Martemyanov}}, \bibinfo {author} {\bibfnamefont {S.~M.}\ \bibnamefont
  {Morozov}}, \bibinfo {author} {\bibfnamefont {M.}~\bibnamefont
  {Muller-Preussker}}, \ and\ \bibinfo {author} {\bibfnamefont {A.~I.}\
  \bibnamefont {Veselov}},\ }\href {\doibase 10.1103/PhysRevD.77.074507}
  {\bibfield  {journal} {\bibinfo  {journal} {Phys. Rev. D}\ }\textbf {\bibinfo
  {volume} {77}},\ \bibinfo {pages} {074507} (\bibinfo {year} {2008})},\
  \Eprint {http://arxiv.org/abs/0708.3335} {arXiv:0708.3335 [hep-lat]}
  \BibitemShut {NoStop}%
\bibitem [{\citenamefont {Bowman}\ \emph {et~al.}(2008)\citenamefont {Bowman},
  \citenamefont {Langfeld}, \citenamefont {Leinweber}, \citenamefont {O'~Cais},
  \citenamefont {Sternbeck}, \citenamefont {von Smekal},\ and\ \citenamefont
  {Williams}}]{Bowman:2008qd}%
  \BibitemOpen
  \bibfield  {author} {\bibinfo {author} {\bibfnamefont {P.~O.}\ \bibnamefont
  {Bowman}}, \bibinfo {author} {\bibfnamefont {K.}~\bibnamefont {Langfeld}},
  \bibinfo {author} {\bibfnamefont {D.~B.}\ \bibnamefont {Leinweber}}, \bibinfo
  {author} {\bibfnamefont {A.}~\bibnamefont {O'~Cais}}, \bibinfo {author}
  {\bibfnamefont {A.}~\bibnamefont {Sternbeck}}, \bibinfo {author}
  {\bibfnamefont {L.}~\bibnamefont {von Smekal}}, \ and\ \bibinfo {author}
  {\bibfnamefont {A.~G.}\ \bibnamefont {Williams}},\ }\href {\doibase
  10.1103/PhysRevD.78.054509} {\bibfield  {journal} {\bibinfo  {journal} {Phys.
  Rev. D}\ }\textbf {\bibinfo {volume} {78}},\ \bibinfo {pages} {054509}
  (\bibinfo {year} {2008})},\ \Eprint {http://arxiv.org/abs/0806.4219}
  {arXiv:0806.4219 [hep-lat]} \BibitemShut {NoStop}%
\bibitem [{\citenamefont {Hollwieser}\ \emph {et~al.}(2008)\citenamefont
  {Hollwieser}, \citenamefont {Faber}, \citenamefont {Greensite}, \citenamefont
  {Heller},\ and\ \citenamefont {Olejnik}}]{Hollwieser:2008tq}%
  \BibitemOpen
  \bibfield  {author} {\bibinfo {author} {\bibfnamefont {R.}~\bibnamefont
  {Hollwieser}}, \bibinfo {author} {\bibfnamefont {M.}~\bibnamefont {Faber}},
  \bibinfo {author} {\bibfnamefont {J.}~\bibnamefont {Greensite}}, \bibinfo
  {author} {\bibfnamefont {U.~M.}\ \bibnamefont {Heller}}, \ and\ \bibinfo
  {author} {\bibfnamefont {S.}~\bibnamefont {Olejnik}},\ }\href {\doibase
  10.1103/PhysRevD.78.054508} {\bibfield  {journal} {\bibinfo  {journal} {Phys.
  Rev. D}\ }\textbf {\bibinfo {volume} {78}},\ \bibinfo {pages} {054508}
  (\bibinfo {year} {2008})},\ \Eprint {http://arxiv.org/abs/0805.1846}
  {arXiv:0805.1846 [hep-lat]} \BibitemShut {NoStop}%
\bibitem [{\citenamefont {O'Cais}\ \emph {et~al.}(2010)\citenamefont {O'Cais},
  \citenamefont {Kamleh}, \citenamefont {Langfeld}, \citenamefont {Lasscock},
  \citenamefont {Leinweber}, \citenamefont {Moran}, \citenamefont {Sternbeck},\
  and\ \citenamefont {von Smekal}}]{OCais:2008kqh}%
  \BibitemOpen
  \bibfield  {author} {\bibinfo {author} {\bibfnamefont {A.}~\bibnamefont
  {O'Cais}}, \bibinfo {author} {\bibfnamefont {W.}~\bibnamefont {Kamleh}},
  \bibinfo {author} {\bibfnamefont {K.}~\bibnamefont {Langfeld}}, \bibinfo
  {author} {\bibfnamefont {B.}~\bibnamefont {Lasscock}}, \bibinfo {author}
  {\bibfnamefont {D.}~\bibnamefont {Leinweber}}, \bibinfo {author}
  {\bibfnamefont {P.}~\bibnamefont {Moran}}, \bibinfo {author} {\bibfnamefont
  {A.}~\bibnamefont {Sternbeck}}, \ and\ \bibinfo {author} {\bibfnamefont
  {L.}~\bibnamefont {von Smekal}},\ }\href {\doibase
  10.1103/PhysRevD.82.114512} {\bibfield  {journal} {\bibinfo  {journal} {Phys.
  Rev. D}\ }\textbf {\bibinfo {volume} {82}},\ \bibinfo {pages} {114512}
  (\bibinfo {year} {2010})},\ \Eprint {http://arxiv.org/abs/0807.0264}
  {arXiv:0807.0264 [hep-lat]} \BibitemShut {NoStop}%
\bibitem [{\citenamefont {Engelhardt}(2011)}]{Engelhardt:2010ft}%
  \BibitemOpen
  \bibfield  {author} {\bibinfo {author} {\bibfnamefont {M.}~\bibnamefont
  {Engelhardt}},\ }\href {\doibase 10.1103/PhysRevD.83.025015} {\bibfield
  {journal} {\bibinfo  {journal} {Phys. Rev.}\ }\textbf {\bibinfo {volume}
  {D83}},\ \bibinfo {pages} {025015} (\bibinfo {year} {2011})},\ \Eprint
  {http://arxiv.org/abs/1008.4953} {arXiv:1008.4953 [hep-lat]} \BibitemShut
  {NoStop}%
\bibitem [{\citenamefont {Bowman}\ \emph {et~al.}(2011)\citenamefont {Bowman},
  \citenamefont {Langfeld}, \citenamefont {Leinweber}, \citenamefont
  {Sternbeck}, \citenamefont {von Smekal},\ and\ \citenamefont
  {Williams}}]{Bowman:2010zr}%
  \BibitemOpen
  \bibfield  {author} {\bibinfo {author} {\bibfnamefont {P.~O.}\ \bibnamefont
  {Bowman}}, \bibinfo {author} {\bibfnamefont {K.}~\bibnamefont {Langfeld}},
  \bibinfo {author} {\bibfnamefont {D.~B.}\ \bibnamefont {Leinweber}}, \bibinfo
  {author} {\bibfnamefont {A.}~\bibnamefont {Sternbeck}}, \bibinfo {author}
  {\bibfnamefont {L.}~\bibnamefont {von Smekal}}, \ and\ \bibinfo {author}
  {\bibfnamefont {A.~G.}\ \bibnamefont {Williams}},\ }\href {\doibase
  10.1103/PhysRevD.84.034501} {\bibfield  {journal} {\bibinfo  {journal} {Phys.
  Rev. D}\ }\textbf {\bibinfo {volume} {84}},\ \bibinfo {pages} {034501}
  (\bibinfo {year} {2011})},\ \Eprint {http://arxiv.org/abs/1010.4624}
  {arXiv:1010.4624 [hep-lat]} \BibitemShut {NoStop}%
\bibitem [{\citenamefont {O'Malley}\ \emph {et~al.}(2012)\citenamefont
  {O'Malley}, \citenamefont {Kamleh}, \citenamefont {Leinweber},\ and\
  \citenamefont {Moran}}]{OMalley:2011aa}%
  \BibitemOpen
  \bibfield  {author} {\bibinfo {author} {\bibfnamefont {E.-A.}\ \bibnamefont
  {O'Malley}}, \bibinfo {author} {\bibfnamefont {W.}~\bibnamefont {Kamleh}},
  \bibinfo {author} {\bibfnamefont {D.}~\bibnamefont {Leinweber}}, \ and\
  \bibinfo {author} {\bibfnamefont {P.}~\bibnamefont {Moran}},\ }\href
  {\doibase 10.1103/PhysRevD.86.054503} {\bibfield  {journal} {\bibinfo
  {journal} {Phys. Rev. D}\ }\textbf {\bibinfo {volume} {86}},\ \bibinfo
  {pages} {054503} (\bibinfo {year} {2012})},\ \Eprint
  {http://arxiv.org/abs/1112.2490} {arXiv:1112.2490 [hep-lat]} \BibitemShut
  {NoStop}%
\bibitem [{\citenamefont {H\"ollwieser}\ \emph {et~al.}(2013)\citenamefont
  {H\"ollwieser}, \citenamefont {Schweigler}, \citenamefont {Faber},\ and\
  \citenamefont {Heller}}]{Hollwieser:2013xja}%
  \BibitemOpen
  \bibfield  {author} {\bibinfo {author} {\bibfnamefont {R.}~\bibnamefont
  {H\"ollwieser}}, \bibinfo {author} {\bibfnamefont {T.}~\bibnamefont
  {Schweigler}}, \bibinfo {author} {\bibfnamefont {M.}~\bibnamefont {Faber}}, \
  and\ \bibinfo {author} {\bibfnamefont {U.~M.}\ \bibnamefont {Heller}},\
  }\href {\doibase 10.1103/PhysRevD.88.114505} {\bibfield  {journal} {\bibinfo
  {journal} {Phys. Rev. D}\ }\textbf {\bibinfo {volume} {88}},\ \bibinfo
  {pages} {114505} (\bibinfo {year} {2013})},\ \Eprint
  {http://arxiv.org/abs/1304.1277} {arXiv:1304.1277 [hep-lat]} \BibitemShut
  {NoStop}%
\bibitem [{\citenamefont {H\"ollwieser}\ \emph {et~al.}(2014)\citenamefont
  {H\"ollwieser}, \citenamefont {Faber}, \citenamefont {Schweigler},\ and\
  \citenamefont {Heller}}]{Hollwieser:2014soz}%
  \BibitemOpen
  \bibfield  {author} {\bibinfo {author} {\bibfnamefont {R.}~\bibnamefont
  {H\"ollwieser}}, \bibinfo {author} {\bibfnamefont {M.}~\bibnamefont {Faber}},
  \bibinfo {author} {\bibfnamefont {T.}~\bibnamefont {Schweigler}}, \ and\
  \bibinfo {author} {\bibfnamefont {U.~M.}\ \bibnamefont {Heller}},\ }\href
  {\doibase 10.22323/1.187.0505} {\bibfield  {journal} {\bibinfo  {journal}
  {PoS}\ }\textbf {\bibinfo {volume} {LATTICE2013}},\ \bibinfo {pages} {505}
  (\bibinfo {year} {2014})},\ \Eprint {http://arxiv.org/abs/1410.2333}
  {arXiv:1410.2333 [hep-lat]} \BibitemShut {NoStop}%
\bibitem [{\citenamefont {Trewartha}\ \emph
  {et~al.}(2015{\natexlab{a}})\citenamefont {Trewartha}, \citenamefont
  {Kamleh},\ and\ \citenamefont {Leinweber}}]{Trewartha:2015ida}%
  \BibitemOpen
  \bibfield  {author} {\bibinfo {author} {\bibfnamefont {D.}~\bibnamefont
  {Trewartha}}, \bibinfo {author} {\bibfnamefont {W.}~\bibnamefont {Kamleh}}, \
  and\ \bibinfo {author} {\bibfnamefont {D.}~\bibnamefont {Leinweber}},\ }\href
  {\doibase 10.1103/PhysRevD.92.074507} {\bibfield  {journal} {\bibinfo
  {journal} {Phys. Rev. D}\ }\textbf {\bibinfo {volume} {92}},\ \bibinfo
  {pages} {074507} (\bibinfo {year} {2015}{\natexlab{a}})},\ \Eprint
  {http://arxiv.org/abs/1509.05518} {arXiv:1509.05518 [hep-lat]} \BibitemShut
  {NoStop}%
\bibitem [{\citenamefont {Trewartha}\ \emph
  {et~al.}(2015{\natexlab{b}})\citenamefont {Trewartha}, \citenamefont
  {Kamleh},\ and\ \citenamefont {Leinweber}}]{Trewartha:2015nna}%
  \BibitemOpen
  \bibfield  {author} {\bibinfo {author} {\bibfnamefont {A.}~\bibnamefont
  {Trewartha}}, \bibinfo {author} {\bibfnamefont {W.}~\bibnamefont {Kamleh}}, \
  and\ \bibinfo {author} {\bibfnamefont {D.}~\bibnamefont {Leinweber}},\ }\href
  {\doibase 10.1016/j.physletb.2015.06.025} {\bibfield  {journal} {\bibinfo
  {journal} {Phys. Lett. B}\ }\textbf {\bibinfo {volume} {747}},\ \bibinfo
  {pages} {373} (\bibinfo {year} {2015}{\natexlab{b}})},\ \Eprint
  {http://arxiv.org/abs/1502.06753} {arXiv:1502.06753 [hep-lat]} \BibitemShut
  {NoStop}%
\bibitem [{\citenamefont {Greensite}(2017)}]{Greensite:2016pfc}%
  \BibitemOpen
  \bibfield  {author} {\bibinfo {author} {\bibfnamefont {J.}~\bibnamefont
  {Greensite}},\ }\href {\doibase 10.1051/epjconf/201713701009} {\bibfield
  {journal} {\bibinfo  {journal} {EPJ Web Conf.}\ }\textbf {\bibinfo {volume}
  {137}},\ \bibinfo {pages} {01009} (\bibinfo {year} {2017})},\ \Eprint
  {http://arxiv.org/abs/1610.06221} {arXiv:1610.06221 [hep-lat]} \BibitemShut
  {NoStop}%
\bibitem [{\citenamefont {Trewartha}\ \emph {et~al.}(2017)\citenamefont
  {Trewartha}, \citenamefont {Kamleh},\ and\ \citenamefont
  {Leinweber}}]{Trewartha:2017ive}%
  \BibitemOpen
  \bibfield  {author} {\bibinfo {author} {\bibfnamefont {A.}~\bibnamefont
  {Trewartha}}, \bibinfo {author} {\bibfnamefont {W.}~\bibnamefont {Kamleh}}, \
  and\ \bibinfo {author} {\bibfnamefont {D.~B.}\ \bibnamefont {Leinweber}},\
  }\href {\doibase 10.1088/1361-6471/aa9443} {\bibfield  {journal} {\bibinfo
  {journal} {J. Phys. G}\ }\textbf {\bibinfo {volume} {44}},\ \bibinfo {pages}
  {125002} (\bibinfo {year} {2017})},\ \Eprint
  {http://arxiv.org/abs/1708.06789} {arXiv:1708.06789 [hep-lat]} \BibitemShut
  {NoStop}%
\bibitem [{\citenamefont {Biddle}\ \emph {et~al.}(2018)\citenamefont {Biddle},
  \citenamefont {Kamleh},\ and\ \citenamefont {Leinweber}}]{Biddle:2018dtc}%
  \BibitemOpen
  \bibfield  {author} {\bibinfo {author} {\bibfnamefont {J.~C.}\ \bibnamefont
  {Biddle}}, \bibinfo {author} {\bibfnamefont {W.}~\bibnamefont {Kamleh}}, \
  and\ \bibinfo {author} {\bibfnamefont {D.~B.}\ \bibnamefont {Leinweber}},\
  }\href {\doibase 10.1103/PhysRevD.98.094504} {\bibfield  {journal} {\bibinfo
  {journal} {Phys. Rev.}\ }\textbf {\bibinfo {volume} {D98}},\ \bibinfo {pages}
  {094504} (\bibinfo {year} {2018})},\ \Eprint
  {http://arxiv.org/abs/1806.04305} {arXiv:1806.04305 [hep-lat]} \BibitemShut
  {NoStop}%
\bibitem [{\citenamefont {Spengler}\ \emph {et~al.}(2018)\citenamefont
  {Spengler}, \citenamefont {Quandt},\ and\ \citenamefont
  {Reinhardt}}]{Spengler:2018dxt}%
  \BibitemOpen
  \bibfield  {author} {\bibinfo {author} {\bibfnamefont {F.}~\bibnamefont
  {Spengler}}, \bibinfo {author} {\bibfnamefont {M.}~\bibnamefont {Quandt}}, \
  and\ \bibinfo {author} {\bibfnamefont {H.}~\bibnamefont {Reinhardt}},\ }\href
  {\doibase 10.1103/PhysRevD.98.094508} {\bibfield  {journal} {\bibinfo
  {journal} {Phys. Rev.}\ }\textbf {\bibinfo {volume} {D98}},\ \bibinfo {pages}
  {094508} (\bibinfo {year} {2018})},\ \Eprint
  {http://arxiv.org/abs/1810.04072} {arXiv:1810.04072 [hep-th]} \BibitemShut
  {NoStop}%
\bibitem [{\citenamefont {Biddle}\ \emph
  {et~al.}(2022{\natexlab{a}})\citenamefont {Biddle}, \citenamefont {Kamleh},\
  and\ \citenamefont {Leinweber}}]{Biddle:2022zgw}%
  \BibitemOpen
  \bibfield  {author} {\bibinfo {author} {\bibfnamefont {J.~C.}\ \bibnamefont
  {Biddle}}, \bibinfo {author} {\bibfnamefont {W.}~\bibnamefont {Kamleh}}, \
  and\ \bibinfo {author} {\bibfnamefont {D.~B.}\ \bibnamefont {Leinweber}},\
  }\href {\doibase 10.1103/PhysRevD.106.054505} {\bibfield  {journal} {\bibinfo
   {journal} {Phys. Rev. D}\ }\textbf {\bibinfo {volume} {106}},\ \bibinfo
  {pages} {054505} (\bibinfo {year} {2022}{\natexlab{a}})},\ \Eprint
  {http://arxiv.org/abs/2206.00844} {arXiv:2206.00844 [hep-lat]} \BibitemShut
  {NoStop}%
\bibitem [{\citenamefont {Biddle}\ \emph
  {et~al.}(2022{\natexlab{b}})\citenamefont {Biddle}, \citenamefont {Kamleh},\
  and\ \citenamefont {Leinweber}}]{Biddle:2022acd}%
  \BibitemOpen
  \bibfield  {author} {\bibinfo {author} {\bibfnamefont {J.~C.}\ \bibnamefont
  {Biddle}}, \bibinfo {author} {\bibfnamefont {W.}~\bibnamefont {Kamleh}}, \
  and\ \bibinfo {author} {\bibfnamefont {D.~B.}\ \bibnamefont {Leinweber}},\
  }\href {\doibase 10.1103/PhysRevD.106.014506} {\bibfield  {journal} {\bibinfo
   {journal} {Phys. Rev. D}\ }\textbf {\bibinfo {volume} {106}},\ \bibinfo
  {pages} {014506} (\bibinfo {year} {2022}{\natexlab{b}})},\ \Eprint
  {http://arxiv.org/abs/2206.02320} {arXiv:2206.02320 [hep-lat]} \BibitemShut
  {NoStop}%
\bibitem [{\citenamefont {Aoki}\ \emph {et~al.}(2009)\citenamefont {Aoki} \emph
  {et~al.}}]{PACS-CS:2008bkb}%
  \BibitemOpen
  \bibfield  {author} {\bibinfo {author} {\bibfnamefont {S.}~\bibnamefont
  {Aoki}} \emph {et~al.},\ }\href {\doibase 10.1103/PhysRevD.79.034503}
  {\bibfield  {journal} {\bibinfo  {journal} {Phys. Rev. D}\ }\textbf {\bibinfo
  {volume} {79}},\ \bibinfo {pages} {034503} (\bibinfo {year} {2009})},\
  \Eprint {http://arxiv.org/abs/0807.1661} {arXiv:0807.1661 [hep-lat]}
  \BibitemShut {NoStop}%
\bibitem [{\citenamefont {Virgili}\ \emph {et~al.}(2022)\citenamefont
  {Virgili}, \citenamefont {Kamleh},\ and\ \citenamefont
  {Leinweber}}]{Virgili:2022ybm}%
  \BibitemOpen
  \bibfield  {author} {\bibinfo {author} {\bibfnamefont {A.}~\bibnamefont
  {Virgili}}, \bibinfo {author} {\bibfnamefont {W.}~\bibnamefont {Kamleh}}, \
  and\ \bibinfo {author} {\bibfnamefont {D.}~\bibnamefont {Leinweber}},\ }\href
  {\doibase 10.1103/PhysRevD.106.014505} {\bibfield  {journal} {\bibinfo
  {journal} {Phys. Rev. D}\ }\textbf {\bibinfo {volume} {106}},\ \bibinfo
  {pages} {014505} (\bibinfo {year} {2022})},\ \Eprint
  {http://arxiv.org/abs/2203.09764} {arXiv:2203.09764 [hep-lat]} \BibitemShut
  {NoStop}%
\bibitem [{\citenamefont {Montero}(1999)}]{Montero:1999by}%
  \BibitemOpen
  \bibfield  {author} {\bibinfo {author} {\bibfnamefont {A.}~\bibnamefont
  {Montero}},\ }\href {\doibase 10.1016/S0370-2693(99)01113-2} {\bibfield
  {journal} {\bibinfo  {journal} {Phys. Lett. B}\ }\textbf {\bibinfo {volume}
  {467}},\ \bibinfo {pages} {106} (\bibinfo {year} {1999})},\ \Eprint
  {http://arxiv.org/abs/hep-lat/9906010} {arXiv:hep-lat/9906010} \BibitemShut
  {NoStop}%
\bibitem [{\citenamefont {Faber}\ \emph {et~al.}(2000)\citenamefont {Faber},
  \citenamefont {Greensite},\ and\ \citenamefont {Olejnik}}]{Faber:1999sq}%
  \BibitemOpen
  \bibfield  {author} {\bibinfo {author} {\bibfnamefont {M.}~\bibnamefont
  {Faber}}, \bibinfo {author} {\bibfnamefont {J.}~\bibnamefont {Greensite}}, \
  and\ \bibinfo {author} {\bibfnamefont {S.}~\bibnamefont {Olejnik}},\ }\href
  {\doibase 10.1016/S0370-2693(00)00013-7} {\bibfield  {journal} {\bibinfo
  {journal} {Phys. Lett. B}\ }\textbf {\bibinfo {volume} {474}},\ \bibinfo
  {pages} {177} (\bibinfo {year} {2000})},\ \Eprint
  {http://arxiv.org/abs/hep-lat/9911006} {arXiv:hep-lat/9911006} \BibitemShut
  {NoStop}%
\bibitem [{\citenamefont {Narayanan}\ and\ \citenamefont
  {Neuberger}(1993{\natexlab{a}})}]{Narayanan:1993zzh}%
  \BibitemOpen
  \bibfield  {author} {\bibinfo {author} {\bibfnamefont {R.}~\bibnamefont
  {Narayanan}}\ and\ \bibinfo {author} {\bibfnamefont {H.}~\bibnamefont
  {Neuberger}},\ }\href {\doibase 10.1016/0370-2693(93)90636-V} {\bibfield
  {journal} {\bibinfo  {journal} {Phys. Lett. B}\ }\textbf {\bibinfo {volume}
  {302}},\ \bibinfo {pages} {62} (\bibinfo {year} {1993}{\natexlab{a}})},\
  \Eprint {http://arxiv.org/abs/hep-lat/9212019} {arXiv:hep-lat/9212019}
  \BibitemShut {NoStop}%
\bibitem [{\citenamefont {Narayanan}\ and\ \citenamefont
  {Neuberger}(1994)}]{Narayanan:1993sk}%
  \BibitemOpen
  \bibfield  {author} {\bibinfo {author} {\bibfnamefont {R.}~\bibnamefont
  {Narayanan}}\ and\ \bibinfo {author} {\bibfnamefont {H.}~\bibnamefont
  {Neuberger}},\ }\href {\doibase 10.1016/0550-3213(94)90393-X} {\bibfield
  {journal} {\bibinfo  {journal} {Nucl. Phys. B}\ }\textbf {\bibinfo {volume}
  {412}},\ \bibinfo {pages} {574} (\bibinfo {year} {1994})},\ \Eprint
  {http://arxiv.org/abs/hep-lat/9307006} {arXiv:hep-lat/9307006} \BibitemShut
  {NoStop}%
\bibitem [{\citenamefont {Narayanan}\ and\ \citenamefont
  {Neuberger}(1993{\natexlab{b}})}]{Narayanan:1993ss}%
  \BibitemOpen
  \bibfield  {author} {\bibinfo {author} {\bibfnamefont {R.}~\bibnamefont
  {Narayanan}}\ and\ \bibinfo {author} {\bibfnamefont {H.}~\bibnamefont
  {Neuberger}},\ }\href {\doibase 10.1103/PhysRevLett.71.3251} {\bibfield
  {journal} {\bibinfo  {journal} {Phys. Rev. Lett.}\ }\textbf {\bibinfo
  {volume} {71}},\ \bibinfo {pages} {3251} (\bibinfo {year}
  {1993}{\natexlab{b}})},\ \Eprint {http://arxiv.org/abs/hep-lat/9308011}
  {arXiv:hep-lat/9308011} \BibitemShut {NoStop}%
\bibitem [{\citenamefont {Narayanan}\ and\ \citenamefont
  {Neuberger}(1995)}]{Narayanan:1994gw}%
  \BibitemOpen
  \bibfield  {author} {\bibinfo {author} {\bibfnamefont {R.}~\bibnamefont
  {Narayanan}}\ and\ \bibinfo {author} {\bibfnamefont {H.}~\bibnamefont
  {Neuberger}},\ }\href {\doibase 10.1016/0550-3213(95)00111-5} {\bibfield
  {journal} {\bibinfo  {journal} {Nucl. Phys. B}\ }\textbf {\bibinfo {volume}
  {443}},\ \bibinfo {pages} {305} (\bibinfo {year} {1995})},\ \Eprint
  {http://arxiv.org/abs/hep-th/9411108} {arXiv:hep-th/9411108} \BibitemShut
  {NoStop}%
\bibitem [{\citenamefont {Neuberger}(1998{\natexlab{a}})}]{Neuberger:1997fp}%
  \BibitemOpen
  \bibfield  {author} {\bibinfo {author} {\bibfnamefont {H.}~\bibnamefont
  {Neuberger}},\ }\href {\doibase 10.1016/S0370-2693(97)01368-3} {\bibfield
  {journal} {\bibinfo  {journal} {Phys. Lett. B}\ }\textbf {\bibinfo {volume}
  {417}},\ \bibinfo {pages} {141} (\bibinfo {year} {1998}{\natexlab{a}})},\
  \Eprint {http://arxiv.org/abs/hep-lat/9707022} {arXiv:hep-lat/9707022}
  \BibitemShut {NoStop}%
\bibitem [{\citenamefont {Kikukawa}\ and\ \citenamefont
  {Neuberger}(1998)}]{Kikukawa:1997qh}%
  \BibitemOpen
  \bibfield  {author} {\bibinfo {author} {\bibfnamefont {Y.}~\bibnamefont
  {Kikukawa}}\ and\ \bibinfo {author} {\bibfnamefont {H.}~\bibnamefont
  {Neuberger}},\ }\href {\doibase 10.1016/S0550-3213(97)00779-7} {\bibfield
  {journal} {\bibinfo  {journal} {Nucl. Phys. B}\ }\textbf {\bibinfo {volume}
  {513}},\ \bibinfo {pages} {735} (\bibinfo {year} {1998})},\ \Eprint
  {http://arxiv.org/abs/hep-lat/9707016} {arXiv:hep-lat/9707016} \BibitemShut
  {NoStop}%
\bibitem [{\citenamefont {Kamleh}\ \emph {et~al.}(2002)\citenamefont {Kamleh},
  \citenamefont {Adams}, \citenamefont {Leinweber},\ and\ \citenamefont
  {Williams}}]{Kamleh:2001ff}%
  \BibitemOpen
  \bibfield  {author} {\bibinfo {author} {\bibfnamefont {W.}~\bibnamefont
  {Kamleh}}, \bibinfo {author} {\bibfnamefont {D.~H.}\ \bibnamefont {Adams}},
  \bibinfo {author} {\bibfnamefont {D.~B.}\ \bibnamefont {Leinweber}}, \ and\
  \bibinfo {author} {\bibfnamefont {A.~G.}\ \bibnamefont {Williams}},\ }\href
  {\doibase 10.1103/PhysRevD.66.014501} {\bibfield  {journal} {\bibinfo
  {journal} {Phys. Rev. D}\ }\textbf {\bibinfo {volume} {66}},\ \bibinfo
  {pages} {014501} (\bibinfo {year} {2002})},\ \Eprint
  {http://arxiv.org/abs/hep-lat/0112041} {arXiv:hep-lat/0112041} \BibitemShut
  {NoStop}%
\bibitem [{\citenamefont {Bietenholz}(2002)}]{Bietenholz:2002ks}%
  \BibitemOpen
  \bibfield  {author} {\bibinfo {author} {\bibfnamefont {W.}~\bibnamefont
  {Bietenholz}},\ }\href {\doibase 10.1016/S0550-3213(02)00789-7} {\bibfield
  {journal} {\bibinfo  {journal} {Nucl. Phys. B}\ }\textbf {\bibinfo {volume}
  {644}},\ \bibinfo {pages} {223} (\bibinfo {year} {2002})},\ \Eprint
  {http://arxiv.org/abs/hep-lat/0204016} {arXiv:hep-lat/0204016} \BibitemShut
  {NoStop}%
\bibitem [{\citenamefont {Kovacs}(2003)}]{Kovacs:2002nz}%
  \BibitemOpen
  \bibfield  {author} {\bibinfo {author} {\bibfnamefont {T.~G.}\ \bibnamefont
  {Kovacs}},\ }\href {\doibase 10.1103/PhysRevD.67.094501} {\bibfield
  {journal} {\bibinfo  {journal} {Phys. Rev. D}\ }\textbf {\bibinfo {volume}
  {67}},\ \bibinfo {pages} {094501} (\bibinfo {year} {2003})},\ \Eprint
  {http://arxiv.org/abs/hep-lat/0209125} {arXiv:hep-lat/0209125} \BibitemShut
  {NoStop}%
\bibitem [{\citenamefont {DeGrand}\ and\ \citenamefont
  {Schaefer}(2005)}]{DeGrand:2004nq}%
  \BibitemOpen
  \bibfield  {author} {\bibinfo {author} {\bibfnamefont {T.~A.}\ \bibnamefont
  {DeGrand}}\ and\ \bibinfo {author} {\bibfnamefont {S.}~\bibnamefont
  {Schaefer}},\ }\href {\doibase 10.1103/PhysRevD.71.034507} {\bibfield
  {journal} {\bibinfo  {journal} {Phys. Rev. D}\ }\textbf {\bibinfo {volume}
  {71}},\ \bibinfo {pages} {034507} (\bibinfo {year} {2005})},\ \Eprint
  {http://arxiv.org/abs/hep-lat/0412005} {arXiv:hep-lat/0412005} \BibitemShut
  {NoStop}%
\bibitem [{\citenamefont {Durr}\ \emph {et~al.}(2006)\citenamefont {Durr},
  \citenamefont {Hoelbling},\ and\ \citenamefont {Wenger}}]{Durr:2005mq}%
  \BibitemOpen
  \bibfield  {author} {\bibinfo {author} {\bibfnamefont {S.}~\bibnamefont
  {Durr}}, \bibinfo {author} {\bibfnamefont {C.}~\bibnamefont {Hoelbling}}, \
  and\ \bibinfo {author} {\bibfnamefont {U.}~\bibnamefont {Wenger}},\ }\href
  {\doibase 10.1016/j.nuclphysbps.2006.01.010} {\bibfield  {journal} {\bibinfo
  {journal} {Nucl. Phys. B Proc. Suppl.}\ }\textbf {\bibinfo {volume} {153}},\
  \bibinfo {pages} {82} (\bibinfo {year} {2006})},\ \Eprint
  {http://arxiv.org/abs/hep-lat/0511046} {arXiv:hep-lat/0511046} \BibitemShut
  {NoStop}%
\bibitem [{\citenamefont {Durr}\ and\ \citenamefont
  {Hoelbling}(2005)}]{Durr:2005ik}%
  \BibitemOpen
  \bibfield  {author} {\bibinfo {author} {\bibfnamefont {S.}~\bibnamefont
  {Durr}}\ and\ \bibinfo {author} {\bibfnamefont {C.}~\bibnamefont
  {Hoelbling}},\ }\href {\doibase 10.1103/PhysRevD.72.071501} {\bibfield
  {journal} {\bibinfo  {journal} {Phys. Rev. D}\ }\textbf {\bibinfo {volume}
  {72}},\ \bibinfo {pages} {071501} (\bibinfo {year} {2005})},\ \Eprint
  {http://arxiv.org/abs/hep-ph/0508085} {arXiv:hep-ph/0508085} \BibitemShut
  {NoStop}%
\bibitem [{\citenamefont {Bietenholz}\ and\ \citenamefont
  {Shcheredin}(2006)}]{Bietenholz:2006fj}%
  \BibitemOpen
  \bibfield  {author} {\bibinfo {author} {\bibfnamefont {W.}~\bibnamefont
  {Bietenholz}}\ and\ \bibinfo {author} {\bibfnamefont {S.}~\bibnamefont
  {Shcheredin}},\ }\href {\doibase 10.1016/j.nuclphysb.2006.07.018} {\bibfield
  {journal} {\bibinfo  {journal} {Nucl. Phys. B}\ }\textbf {\bibinfo {volume}
  {754}},\ \bibinfo {pages} {17} (\bibinfo {year} {2006})},\ \Eprint
  {http://arxiv.org/abs/hep-lat/0605013} {arXiv:hep-lat/0605013} \BibitemShut
  {NoStop}%
\bibitem [{\citenamefont {Zanotti}\ \emph {et~al.}(2002)\citenamefont
  {Zanotti}, \citenamefont {Bilson-Thompson}, \citenamefont {Bonnet},
  \citenamefont {Coddington}, \citenamefont {Leinweber}, \citenamefont
  {Williams}, \citenamefont {Zhang}, \citenamefont {Melnitchouk},\ and\
  \citenamefont {Lee}}]{Zanotti:2001yb}%
  \BibitemOpen
  \bibfield  {author} {\bibinfo {author} {\bibfnamefont {J.~M.}\ \bibnamefont
  {Zanotti}}, \bibinfo {author} {\bibfnamefont {S.~O.}\ \bibnamefont
  {Bilson-Thompson}}, \bibinfo {author} {\bibfnamefont {F.~D.~R.}\ \bibnamefont
  {Bonnet}}, \bibinfo {author} {\bibfnamefont {P.~D.}\ \bibnamefont
  {Coddington}}, \bibinfo {author} {\bibfnamefont {D.~B.}\ \bibnamefont
  {Leinweber}}, \bibinfo {author} {\bibfnamefont {A.~G.}\ \bibnamefont
  {Williams}}, \bibinfo {author} {\bibfnamefont {J.~B.}\ \bibnamefont {Zhang}},
  \bibinfo {author} {\bibfnamefont {W.}~\bibnamefont {Melnitchouk}}, \ and\
  \bibinfo {author} {\bibfnamefont {F.~X.}\ \bibnamefont {Lee}},\ }\href
  {\doibase 10.1103/PhysRevD.65.074507} {\bibfield  {journal} {\bibinfo
  {journal} {Phys. Rev. D}\ }\textbf {\bibinfo {volume} {65}},\ \bibinfo
  {pages} {074507} (\bibinfo {year} {2002})},\ \Eprint
  {http://arxiv.org/abs/hep-lat/0110216} {arXiv:hep-lat/0110216} \BibitemShut
  {NoStop}%
\bibitem [{\citenamefont {Kamleh}\ \emph {et~al.}(2004)\citenamefont {Kamleh},
  \citenamefont {Leinweber},\ and\ \citenamefont {Williams}}]{Kamleh:2004xk}%
  \BibitemOpen
  \bibfield  {author} {\bibinfo {author} {\bibfnamefont {W.}~\bibnamefont
  {Kamleh}}, \bibinfo {author} {\bibfnamefont {D.~B.}\ \bibnamefont
  {Leinweber}}, \ and\ \bibinfo {author} {\bibfnamefont {A.~G.}\ \bibnamefont
  {Williams}},\ }\href {\doibase 10.1103/PhysRevD.70.014502} {\bibfield
  {journal} {\bibinfo  {journal} {Phys. Rev. D}\ }\textbf {\bibinfo {volume}
  {70}},\ \bibinfo {pages} {014502} (\bibinfo {year} {2004})},\ \Eprint
  {http://arxiv.org/abs/hep-lat/0403019} {arXiv:hep-lat/0403019} \BibitemShut
  {NoStop}%
\bibitem [{\citenamefont {Kamleh}\ \emph {et~al.}(2005)\citenamefont {Kamleh},
  \citenamefont {Bowman}, \citenamefont {Leinweber}, \citenamefont {Williams},\
  and\ \citenamefont {Zhang}}]{Kamleh:2004aw}%
  \BibitemOpen
  \bibfield  {author} {\bibinfo {author} {\bibfnamefont {W.}~\bibnamefont
  {Kamleh}}, \bibinfo {author} {\bibfnamefont {P.~O.}\ \bibnamefont {Bowman}},
  \bibinfo {author} {\bibfnamefont {D.~B.}\ \bibnamefont {Leinweber}}, \bibinfo
  {author} {\bibfnamefont {A.~G.}\ \bibnamefont {Williams}}, \ and\ \bibinfo
  {author} {\bibfnamefont {J.}~\bibnamefont {Zhang}},\ }\href {\doibase
  10.1103/PhysRevD.71.094507} {\bibfield  {journal} {\bibinfo  {journal} {Phys.
  Rev. D}\ }\textbf {\bibinfo {volume} {71}},\ \bibinfo {pages} {094507}
  (\bibinfo {year} {2005})},\ \Eprint {http://arxiv.org/abs/hep-lat/0412022}
  {arXiv:hep-lat/0412022} \BibitemShut {NoStop}%
\bibitem [{\citenamefont {Virgili}\ \emph {et~al.}(2023)\citenamefont
  {Virgili}, \citenamefont {Kamleh},\ and\ \citenamefont
  {Leinweber}}]{Virgili:2022wfx}%
  \BibitemOpen
  \bibfield  {author} {\bibinfo {author} {\bibfnamefont {A.}~\bibnamefont
  {Virgili}}, \bibinfo {author} {\bibfnamefont {W.}~\bibnamefont {Kamleh}}, \
  and\ \bibinfo {author} {\bibfnamefont {D.~B.}\ \bibnamefont {Leinweber}},\
  }\href {\doibase 10.1016/j.physletb.2023.137865} {\bibfield  {journal}
  {\bibinfo  {journal} {Phys. Lett. B}\ }\textbf {\bibinfo {volume} {840}},\
  \bibinfo {pages} {137865} (\bibinfo {year} {2023})},\ \Eprint
  {http://arxiv.org/abs/2209.14864} {arXiv:2209.14864 [hep-lat]} \BibitemShut
  {NoStop}%
\bibitem [{\citenamefont {Lepage}\ and\ \citenamefont
  {Mackenzie}(1993)}]{LePage:1992xa}%
  \BibitemOpen
  \bibfield  {author} {\bibinfo {author} {\bibfnamefont {G.~P.}\ \bibnamefont
  {Lepage}}\ and\ \bibinfo {author} {\bibfnamefont {P.~B.}\ \bibnamefont
  {Mackenzie}},\ }\href {\doibase 10.1103/PhysRevD.48.2250} {\bibfield
  {journal} {\bibinfo  {journal} {Phys. Rev. D}\ }\textbf {\bibinfo {volume}
  {48}},\ \bibinfo {pages} {2250} (\bibinfo {year} {1993})},\ \Eprint
  {http://arxiv.org/abs/hep-lat/9209022} {arXiv:hep-lat/9209022} \BibitemShut
  {NoStop}%
\bibitem [{\citenamefont {Morningstar}\ and\ \citenamefont
  {Peardon}(2004)}]{Morningstar:2003gk}%
  \BibitemOpen
  \bibfield  {author} {\bibinfo {author} {\bibfnamefont {C.}~\bibnamefont
  {Morningstar}}\ and\ \bibinfo {author} {\bibfnamefont {M.}~\bibnamefont
  {Peardon}},\ }\href {\doibase 10.1103/physrevd.69.054501} {\bibfield
  {journal} {\bibinfo  {journal} {Phys. Rev. D}\ }\textbf {\bibinfo {volume}
  {69}},\ \bibinfo {pages} {054501} (\bibinfo {year} {2004})},\ \Eprint
  {http://arxiv.org/abs/hep-lat/0311018} {arXiv:hep-lat/0311018} \BibitemShut
  {NoStop}%
\bibitem [{\citenamefont {Sheikholeslami}\ and\ \citenamefont
  {Wohlert}(1985)}]{Sheikholeslami:1985ij}%
  \BibitemOpen
  \bibfield  {author} {\bibinfo {author} {\bibfnamefont {B.}~\bibnamefont
  {Sheikholeslami}}\ and\ \bibinfo {author} {\bibfnamefont {R.}~\bibnamefont
  {Wohlert}},\ }\href {\doibase 10.1016/0550-3213(85)90002-1} {\bibfield
  {journal} {\bibinfo  {journal} {Nucl. Phys. B}\ }\textbf {\bibinfo {volume}
  {259}},\ \bibinfo {pages} {572} (\bibinfo {year} {1985})}\BibitemShut
  {NoStop}%
\bibitem [{\citenamefont {Neuberger}(1998{\natexlab{b}})}]{Neuberger:1997bg}%
  \BibitemOpen
  \bibfield  {author} {\bibinfo {author} {\bibfnamefont {H.}~\bibnamefont
  {Neuberger}},\ }\href {\doibase 10.1103/PhysRevD.57.5417} {\bibfield
  {journal} {\bibinfo  {journal} {Phys. Rev. D}\ }\textbf {\bibinfo {volume}
  {57}},\ \bibinfo {pages} {5417} (\bibinfo {year} {1998}{\natexlab{b}})},\
  \Eprint {http://arxiv.org/abs/hep-lat/9710089} {arXiv:hep-lat/9710089}
  \BibitemShut {NoStop}%
\bibitem [{\citenamefont {Edwards}\ \emph {et~al.}(1999)\citenamefont
  {Edwards}, \citenamefont {Heller},\ and\ \citenamefont
  {Narayanan}}]{Edwards:1998wx}%
  \BibitemOpen
  \bibfield  {author} {\bibinfo {author} {\bibfnamefont {R.~G.}\ \bibnamefont
  {Edwards}}, \bibinfo {author} {\bibfnamefont {U.~M.}\ \bibnamefont {Heller}},
  \ and\ \bibinfo {author} {\bibfnamefont {R.}~\bibnamefont {Narayanan}},\
  }\href {\doibase 10.1103/PhysRevD.59.094510} {\bibfield  {journal} {\bibinfo
  {journal} {Phys. Rev. D}\ }\textbf {\bibinfo {volume} {59}},\ \bibinfo
  {pages} {094510} (\bibinfo {year} {1999})},\ \Eprint
  {http://arxiv.org/abs/hep-lat/9811030} {arXiv:hep-lat/9811030} \BibitemShut
  {NoStop}%
\bibitem [{\citenamefont {Bonnet}\ \emph {et~al.}(1999)\citenamefont {Bonnet},
  \citenamefont {Bowman}, \citenamefont {Leinweber}, \citenamefont {Williams},\
  and\ \citenamefont {Richards}}]{Bonnet:1999mj}%
  \BibitemOpen
  \bibfield  {author} {\bibinfo {author} {\bibfnamefont {F.~D.~R.}\
  \bibnamefont {Bonnet}}, \bibinfo {author} {\bibfnamefont {P.~O.}\
  \bibnamefont {Bowman}}, \bibinfo {author} {\bibfnamefont {D.~B.}\
  \bibnamefont {Leinweber}}, \bibinfo {author} {\bibfnamefont {A.~G.}\
  \bibnamefont {Williams}}, \ and\ \bibinfo {author} {\bibfnamefont {D.~G.}\
  \bibnamefont {Richards}},\ }\href {\doibase 10.1071/PH99047} {\bibfield
  {journal} {\bibinfo  {journal} {Austral. J. Phys.}\ }\textbf {\bibinfo
  {volume} {52}},\ \bibinfo {pages} {939} (\bibinfo {year} {1999})},\ \Eprint
  {http://arxiv.org/abs/hep-lat/9905006} {arXiv:hep-lat/9905006} \BibitemShut
  {NoStop}%
\bibitem [{\citenamefont {Hudspith}(2015)}]{Hudspith:2014oja}%
  \BibitemOpen
  \bibfield  {author} {\bibinfo {author} {\bibfnamefont {R.~J.}\ \bibnamefont
  {Hudspith}} (\bibinfo {collaboration} {RBC, UKQCD}),\ }\href {\doibase
  10.1016/j.cpc.2014.10.017} {\bibfield  {journal} {\bibinfo  {journal}
  {Comput. Phys. Commun.}\ }\textbf {\bibinfo {volume} {187}},\ \bibinfo
  {pages} {115} (\bibinfo {year} {2015})},\ \Eprint
  {http://arxiv.org/abs/1405.5812} {arXiv:1405.5812 [hep-lat]} \BibitemShut
  {NoStop}%
\bibitem [{\citenamefont {Chiu}\ \emph {et~al.}(2002)\citenamefont {Chiu},
  \citenamefont {Hsieh}, \citenamefont {Huang},\ and\ \citenamefont
  {Huang}}]{Chiu:2002eh}%
  \BibitemOpen
  \bibfield  {author} {\bibinfo {author} {\bibfnamefont {T.-W.}\ \bibnamefont
  {Chiu}}, \bibinfo {author} {\bibfnamefont {T.-H.}\ \bibnamefont {Hsieh}},
  \bibinfo {author} {\bibfnamefont {C.-H.}\ \bibnamefont {Huang}}, \ and\
  \bibinfo {author} {\bibfnamefont {T.-R.}\ \bibnamefont {Huang}},\ }\href
  {\doibase 10.1103/PhysRevD.66.114502} {\bibfield  {journal} {\bibinfo
  {journal} {Phys. Rev. D}\ }\textbf {\bibinfo {volume} {66}},\ \bibinfo
  {pages} {114502} (\bibinfo {year} {2002})},\ \Eprint
  {http://arxiv.org/abs/hep-lat/0206007} {arXiv:hep-lat/0206007} \BibitemShut
  {NoStop}%
\bibitem [{\citenamefont {Leinweber}\ \emph {et~al.}(1998)\citenamefont
  {Leinweber}, \citenamefont {Skullerud}, \citenamefont {Williams},\ and\
  \citenamefont {Parrinello}}]{Leinweber:1998im}%
  \BibitemOpen
  \bibfield  {author} {\bibinfo {author} {\bibfnamefont {D.~B.}\ \bibnamefont
  {Leinweber}}, \bibinfo {author} {\bibfnamefont {J.~I.}\ \bibnamefont
  {Skullerud}}, \bibinfo {author} {\bibfnamefont {A.~G.}\ \bibnamefont
  {Williams}}, \ and\ \bibinfo {author} {\bibfnamefont {C.}~\bibnamefont
  {Parrinello}} (\bibinfo {collaboration} {UKQCD}),\ }\href {\doibase
  10.1103/PhysRevD.58.031501} {\bibfield  {journal} {\bibinfo  {journal} {Phys.
  Rev. D}\ }\textbf {\bibinfo {volume} {58}},\ \bibinfo {pages} {031501}
  (\bibinfo {year} {1998})},\ \Eprint {http://arxiv.org/abs/hep-lat/9803015}
  {arXiv:hep-lat/9803015} \BibitemShut {NoStop}%
\bibitem [{\citenamefont {Leinweber}\ \emph {et~al.}(1999)\citenamefont
  {Leinweber}, \citenamefont {Skullerud}, \citenamefont {Williams},\ and\
  \citenamefont {Parrinello}}]{Leinweber:1998uu}%
  \BibitemOpen
  \bibfield  {author} {\bibinfo {author} {\bibfnamefont {D.~B.}\ \bibnamefont
  {Leinweber}}, \bibinfo {author} {\bibfnamefont {J.~I.}\ \bibnamefont
  {Skullerud}}, \bibinfo {author} {\bibfnamefont {A.~G.}\ \bibnamefont
  {Williams}}, \ and\ \bibinfo {author} {\bibfnamefont {C.}~\bibnamefont
  {Parrinello}} (\bibinfo {collaboration} {UKQCD}),\ }\href {\doibase
  10.1103/PhysRevD.60.094507} {\bibfield  {journal} {\bibinfo  {journal} {Phys.
  Rev. D}\ }\textbf {\bibinfo {volume} {60}},\ \bibinfo {pages} {094507}
  (\bibinfo {year} {1999})},\ \bibinfo {note} {[Erratum: Phys.Rev.D 61, 079901
  (2000)]},\ \Eprint {http://arxiv.org/abs/hep-lat/9811027}
  {arXiv:hep-lat/9811027} \BibitemShut {NoStop}%
\bibitem [{\citenamefont {Bilson-Thompson}\ \emph {et~al.}(2003)\citenamefont
  {Bilson-Thompson}, \citenamefont {Leinweber},\ and\ \citenamefont
  {Williams}}]{Bilson-Thompson:2002xlt}%
  \BibitemOpen
  \bibfield  {author} {\bibinfo {author} {\bibfnamefont {S.~O.}\ \bibnamefont
  {Bilson-Thompson}}, \bibinfo {author} {\bibfnamefont {D.~B.}\ \bibnamefont
  {Leinweber}}, \ and\ \bibinfo {author} {\bibfnamefont {A.~G.}\ \bibnamefont
  {Williams}},\ }\href {\doibase 10.1016/S0003-4916(03)00009-5} {\bibfield
  {journal} {\bibinfo  {journal} {Annals Phys.}\ }\textbf {\bibinfo {volume}
  {304}},\ \bibinfo {pages} {1} (\bibinfo {year} {2003})},\ \Eprint
  {http://arxiv.org/abs/hep-lat/0203008} {arXiv:hep-lat/0203008} \BibitemShut
  {NoStop}%
\bibitem [{\citenamefont {Moran}\ and\ \citenamefont
  {Leinweber}(2008)}]{Moran:2008qd}%
  \BibitemOpen
  \bibfield  {author} {\bibinfo {author} {\bibfnamefont {P.~J.}\ \bibnamefont
  {Moran}}\ and\ \bibinfo {author} {\bibfnamefont {D.~B.}\ \bibnamefont
  {Leinweber}},\ }\href {\doibase 10.1103/PhysRevD.78.054506} {\bibfield
  {journal} {\bibinfo  {journal} {Phys. Rev. D}\ }\textbf {\bibinfo {volume}
  {78}},\ \bibinfo {pages} {054506} (\bibinfo {year} {2008})},\ \Eprint
  {http://arxiv.org/abs/0801.2016} {arXiv:0801.2016 [hep-lat]} \BibitemShut
  {NoStop}%
\bibitem [{\citenamefont {Kamleh}(2023)}]{Kamleh:2022nqr}%
  \BibitemOpen
  \bibfield  {author} {\bibinfo {author} {\bibfnamefont {W.}~\bibnamefont
  {Kamleh}},\ }\href {\doibase 10.22323/1.430.0339} {\bibfield  {journal}
  {\bibinfo  {journal} {PoS}\ }\textbf {\bibinfo {volume} {LATTICE2022}},\
  \bibinfo {pages} {339} (\bibinfo {year} {2023})},\ \Eprint
  {http://arxiv.org/abs/2302.00850} {arXiv:2302.00850 [hep-lat]} \BibitemShut
  {NoStop}%
\end{thebibliography}

%

\end{document}